\documentclass[usenatbib,usegraphicx,letterpaper]{mn2e}

\usepackage{lmodern}

\usepackage{amsmath}
\usepackage{amstext}
\usepackage{amssymb}
\usepackage{yfonts}

\usepackage{epsfig}
\usepackage{graphicx}
\usepackage{color}

\usepackage[lowtilde]{url}
\usepackage{hyperref}

\newcommand{\apjl}{ApJL~}
\newcommand{\apjs}{ApJS~}

\newcommand{\beq}{\begin{equation}}
\newcommand{\eeq}{\end{equation}}
\newcommand{\beqray}{\begin{eqnarray}}
\newcommand{\eeqray}{\end{eqnarray}}

\newcommand{\ben}{\begin{enumerate}}
\newcommand{\een}{\end{enumerate}}
\newcommand{\bit}{\begin{itemize}}
\newcommand{\eit}{\end{itemize}}

\newcommand{\rhalf}{R_{1/2}}

\newcommand{\mstar}{M_{\ast}}

\newcommand{\fsat}{F_{\rm sat}}

\newcommand{\mvir}{M_{\rm vir}}

\newcommand{\mpeak}{M_{\rm peak}}
\newcommand{\zpeak}{z_{M_{\rm peak}}}
\newcommand{\mhalo}{M_{\rm halo}}
\newcommand{\mhost}{M_{\rm host}}
\newcommand{\rvir}{R_{\rm vir}}
\newcommand{\rmpeak}{R_{\rm M_{peak}}}

\newcommand{\rproj}{r_{\rm p}}
\newcommand{\wproj}{w_{\rm p}}
\newcommand{\wplarge}{w_{\rm p}^{\rm large}}
\newcommand{\wpsmall}{w_{\rm p}^{\rm small}}
\newcommand{\wpall}{w_{\rm p}^{\rm all}}
\newcommand{\mean}[2]{\langle{#1}\vert{#2}\rangle}
\newcommand{\median}[2]{\langle{#1}\vert{#2}\rangle_{\rm median}}

\newcommand{\kpc}{{\rm kpc}}
\newcommand{\mpc}{{\rm Mpc}}
\newcommand{\msun}{M_\odot}
\newcommand{\kms}{{\rm km/s}}

\usepackage{epsfig}  \usepackage{graphicx}   \usepackage{rotating}

\begin{document}

\title[The Relative Sizes of Centrals and Satellites]
{Clustering Constraints on the Relative Sizes of Central and Satellite Galaxies}

\author[Hearin, Behroozi, Kravtsov \& Moster]{
Andrew Hearin$^{1}$, Peter Behroozi$^{2}$, Andrey Kravtsov$^{3}$, Benjamin Moster$^{4,5}$\\
$^{1}$Argonne National Laboratory, Argonne, IL, USA 60439, USA\\
$^{2}$Department of Physics, University of Arizona, 1118 E 4th St, Tucson, AZ 85721 USA\\
$^{3}$Department of Astronomy \& Astrophysics, and Kavli Institute for Cosmological Physics, The University of Chicago, Chicago IL 60637\\
$^{4}$Universit{\"a}ts-Sternwarte, Ludwig-Maximilians-Universit{\"a}t M{\"u}nchen, Scheinerstr. 1, 81679 M{\"u}nchen, Germany\\
$^5$Max-Planck Institut f\"ur Astrophysik, Karl-Schwarzschild Stra\ss e 1, 85748 Garching, Germany
}

\maketitle

\begin{abstract}
We empirically constrain how galaxy size relates to halo virial radius using new measurements of the size- and stellar mass-dependent clustering of galaxies in the Sloan Digital Sky Survey.  We find that small galaxies cluster much more strongly than large galaxies of the same stellar mass. 
The magnitude of this clustering difference increases on small scales, and decreases with increasing stellar mass. 
Using {\tt Halotools} to forward model the observations, we test an empirical model in which present-day galaxy size is 
proportional to the size of the virial radius at the time the halo reached its maximum mass. This simple model reproduces the 
observed size-dependence of galaxy clustering in striking detail. 
The success of this model provides strong support for the conclusion that satellite galaxies have smaller sizes relative to central 
galaxies of the same halo mass. 
Our findings indicate that satellite size is set prior to the time of infall, and that a remarkably simple, linear size--virial radius relation 
emerges from the complex physics regulating galaxy size. 
We make quantitative predictions for future measurements of galaxy-galaxy lensing, including dependence upon size, scale, 
and stellar mass, and provide a scaling relation of the ratio of mean sizes of satellites and central galaxies as a function of their halo mass 
that can be used to calibrate hydrodynamical simulations and semi-analytic models. 
\end{abstract}

\section{Introduction}
\label{sec:intro}

Many properties of observed galaxies exhibit remarkably tight scaling relations. Galaxy size, typically quantified by a half-mass or half-light radius, $\rhalf,$ has well-measured scaling with galaxy mass $\mstar$ in the local Universe \citep{shen_etal03,huang_etal13,lange_etal15,zhang_yang17}, and at higher redshift \citep{trujillo_etal04,huertas_company_etal13a,vanderwel_etal14,kawamata_etal15,shibuya_etal15,huang_etal17}.

Observational constraints on models for galaxy size typically come from one-point function measurements such as $\mean{\rhalf}{\mstar}$ and $\sigma(\rhalf\vert\mstar),$ e.g.,  \citet{khochfar_silk06,lang_etal14,desmond_etal17,bottrell_etal17b,hou_etal17,somerville_etal17}, or otherwise from catalogs of galaxy groups \citep{weinmann_etal08,guo_etal09,huertas_company_etal13b,spindler_wake17}. More recently, the connection between a galaxy's size and the virial radius of its parent halo has been
explored by converting stellar mass to halo mass and virial radius, using abundance matching \citep{kravtsov_etal04,tasitsiomi_etal04,vale_ostriker04,vale_ostriker06,conroy_etal06}.  Such analyses have indicated that galaxy size is, on average, proportional to host halo virial radius, both at $z=0$ \citep{kravtsov13}, and higher redshifts \citep{huang_etal17,somerville_etal17}.

Observations of two-point galaxy clustering, $\wproj(\rproj),$ have historically been used to place tight constraints on many features of the galaxy--halo connection, such as its $\mstar-$dependence \citep{moster_etal10,leauthaud_etal11,reddick_etal13,skibba_etal15}, dependence on luminosity \citep{kravtsov_etal04, tasitsiomi_etal04,vale_ostriker04,vale_ostriker06,tinker_etal05,cacciato_etal13}, broadband color \citep{coil_etal08,zehavi_etal11,guo_etal11,hearin_watson13}, and star-formation rate \citep{wang_etal07,tinker_etal13,watson_etal14}. In the present work, we exploit the constraining power of $\wproj(\rproj)$ to test empirical models connecting galaxy stellar mass and size to the properties of the galaxy's parent dark matter halo.

The approach we take is to forward model the $\rhalf-$halo connection: we generate Monte Carlo realizations of models using a high-resolution cosmological $N$-body simulation, and then make synthetic measurements in a manner that closely mimics the measurements we make on observed galaxies. We describe the dataset and simulations we use in \S\ref{sec:data}. Our forward-modeling methods are described in \S\ref{sec:model}, with supplementary material in the Appendix. All of our primary results appear in \S\ref{sec:results}, and their theoretical interpretation in \S\ref{sec:interpretation}. We contextualize our findings in terms of previous work in \S\ref{sec:previous_work}, and discuss future directions in \S\ref{sec:future}. We conclude the paper with a summary of our principal results in \S\ref{sec:conclusion}.

\begin{figure*}
\centering
\includegraphics[width=9cm]{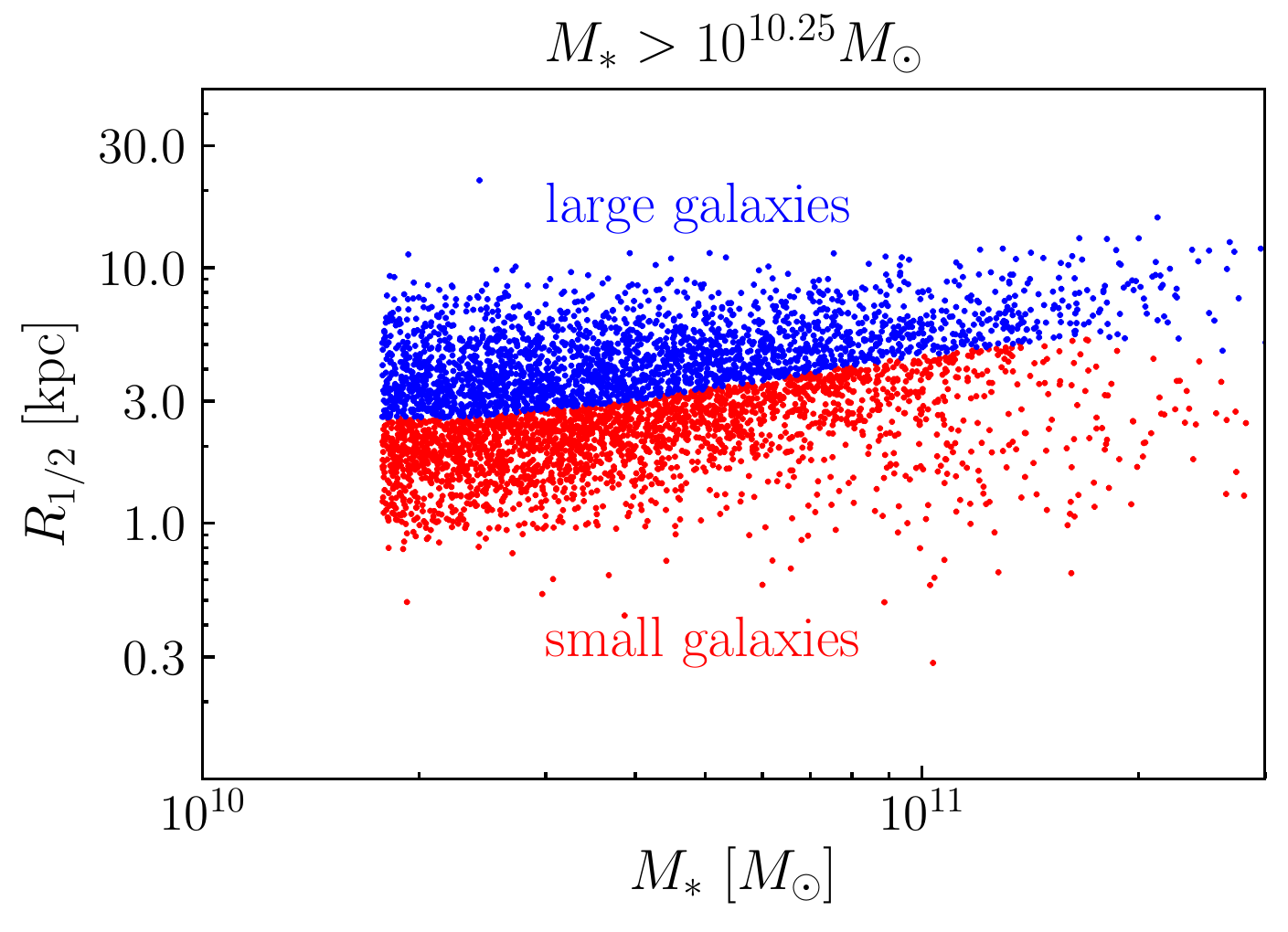}
\includegraphics[width=\textwidth]{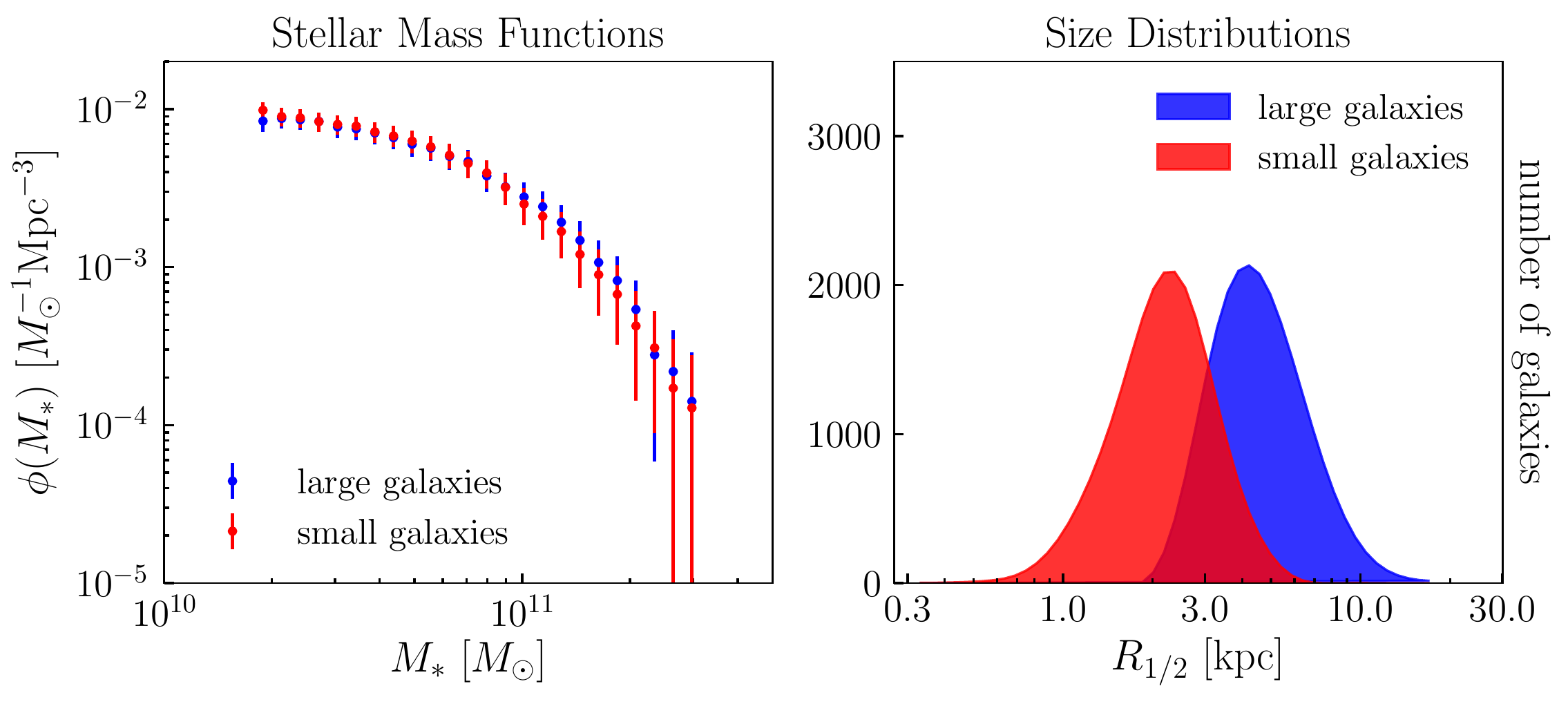}
\caption{
{\bf Definition of ``small" and ``large" galaxies.} For a volume-limited SDSS galaxy sample defined by $M_{\ast}>10^{10.25}M_{\odot},$ the {\em top panel} visually demonstrates how we classify galaxies into ``small" and ``large" subsamples in an $\mstar$-dependent manner. As described in detail in \S\ref{subsec:sizedef}, we compute the median value $\median{\rhalf}{\mstar}$ at the stellar mass of each galaxy in the sample, and divide the sample into two based on this $\mstar-$dependent cut. The {\em bottom left panel} compares the stellar mass functions $\phi(\mstar)$ of the two samples, confirming that our method for separating ``small" from ``large" galaxies yields subsamples with statistically indistinguishable $\phi(\mstar)$. The {\em bottom right panel} shows histograms of $\rhalf$ for the two subsamples, which partially overlap due to the stellar mass range spanned by the sample.
}
\label{fig:sizedefinition}
\end{figure*}

\section{Data and Simulations}
\label{sec:data}

The starting point for our galaxy samples comes from Data Release 10 of the Sloan Digital Sky Survey \citep[SDSS,][]{ahn_etal14}, with stellar mass measurements taken from the MPA-JHU catalog \citep{kauffmann_etal03,brinchmann_etal04}. Discarding galaxies with redshift $z<0.02,$ we define volume-limited, $\mstar-$threshold samples by applying a maximum-redshift cut $z_{\rm max}$ appropriate for the $\mstar$ threshold; we use the same $z_{\rm max}$ cut shown in Figure 2 of \citet{behroozi_etal15}.

We supplement the MPA-JHU catalog with measurements of half-light radius, $\rhalf,$ derived from galaxy profile decompositions provided by \citet{meert_etal15}. The \citet{meert_etal15} catalog is based on Data Release 7 of SDSS \citep{abazajian_etal09}, with improvements to the photometry pipeline and light profile fitting method and, especially, in background subtraction \citep{vikram_etal10,bernardi_etal13,bernardi_etal14,meert_etal13}. In the version of this catalog that we use,\footnote{Our $\mstar$ and $\rhalf$ measurements are derived from different photometry pipelines; this is driven by our desire for consistency with the \citet{behroozi_etal15} completeness cuts used in our clustering measurements. We note that we have repeated our analysis for stellar mass measurements based on \citet{meert_etal15}, finding only minor quantitative, and no qualitative changes to our results.} two-dimensional $r-$band profiles were fit with a two-component de Vaucouleurs + exponential profile to determine the projected half-light radius of total $r-$band luminosity, $\rhalf.$

As the bedrock of our modeling, we use the publicly available\footnote{\url{http://www.slac.stanford.edu/~behroozi/BPlanck\_Hlists}} catalog of {\tt Rockstar} subhalos identified at $z=0$ in the Bolshoi-Planck simulation \citep{klypin_etal11,klypin_etal16,behroozi12_rockstar,riebe_etal13,behroozi_etal12b,rodriguez_puebla16_bolplanck}. As described in \S\ref{subsec:sham}, we use traditional abundance matching to connect stellar mass $\mstar$ with masses of halos and subhalos. To address issues related to subhalo incompleteness \citep{guo_white13,campbell_etal17}, we supplement the Bolshoi-Planck subhalo catalog with subhalos that have been disrupted and no longer appear in the standard catalog as $z=0$ surviving subhalos. We describe our treatment of these ``orphan" galaxies in the Appendix, where we also show results that exclude all orphan galaxies. 

For our SDSS galaxy samples, we calculate two-point clustering $\wproj$ using line-of-sight projection of $\pi_{\rm max}=20~\mpc$ using the {\tt correl} program in {\tt UniverseMachine} (Behroozi et al. 2017, in prep). The {\tt correl} code uses the \citet{landy_szalay93} estimator ($DD - 2DR + RR$) to compute the redshift-space correlation function $\xi(\rproj, \pi)$, which is then integrated over $\vert\pi\vert < 20$ Mpc to compute the projected correlation function $\wproj(\rproj)$.  The code uses $10^6$ randoms drawn from the same mask region with uniform volume distribution to compute $DR$, and $RR$ is computed via Monte Carlo integration. Errors are estimated by jackknife resampling, giving us an estimate for the lower bound of samples with volumes $V_{\rm eff}\lesssim 0.3\,{\rm Gpc}^3.$

For mock galaxies, we compute $\wproj$ using the {\tt mock\_observables.wp} function in {\tt Halotools} \citep{hearin_etal16}, which is a python implementation of the algorithm in the {\tt Corrfunc} C library \citep{sinha_etal17}. We also use {\tt Halotools} to compute the galaxy-galaxy lensing signal, $\Delta\Sigma,$ using the {\tt mock\_observables.delta\_sigma} function.

All numerical values of $\rhalf$ will be quoted in physical $\kpc,$ and all values of $\mstar$ and $\mhalo$ in $\msun,$ assuming $H_0=67.8~\kms\equiv100h~\kms,$ the best-fit value from \citet{planck15}. To scale stellar masses to ``$h=1$ units" \citep{croton13}, our numerically quoted values for $\mstar$ should be multiplied by $h^2,$ while our halo masses and distances should be multiplied by $h.$

\subsection{Classifying large vs. small galaxies}
\label{subsec:sizedef}

Galaxy clustering has well-known dependence upon $\mstar$, which is not the focus of this work.  We thus wish to remove this dependence to focus solely on the  $\wproj(\rproj)$ dependence
on $\rhalf.$ To do so, we determine the value $\median{\rhalf}{\mstar}$ by computing a sliding median of $\rhalf$ for galaxies sorted by $\mstar,$ calculated using a symmetric window of width $N_{\rm gal}=1000,$ keeping this window fixed for the first and last $500$ galaxies. Each galaxy is categorized as either ``large" or ``small" according to whether it is above or below the median value appropriate for its stellar mass. We note that this is analogous to the common convention for studying the properties of ``red" vs. ``blue" galaxies, in which the two subsamples are divided by a $\mstar-$ or luminosity-dependent green valley cut \citep[e.g.,][]{vdB_etal08,zehavi_etal11}, only here the $\rhalf$ distribution is uni-modal, not bi-modal.

Using this technique, for any $\mstar-$threshold sample, the stellar mass function of the ``large" and ``small" subsamples are identical. We illustrate this for the particular case of $\mstar>10^{10.25}\msun$ in bottom left panel of Figure~\ref{fig:sizedefinition}, which shows stellar mass functions for the two subsamples. The bottom right panel of Figure~\ref{fig:sizedefinition} compares histograms of the two size distributions of these subsamples, which partially overlap due to the variation in $\median{\rhalf}{\mstar}$ across the $\mstar-$ range of the threshold sample.

In addition to stellar mass, $\wproj(\rproj)$ depends on galaxy color \citep[e.g.,][]{zehavi_etal11}. Thus, we will also consider the 
dependence of clustering on galaxy size while controlling for color. Observations of clustering exhibited by subsamples split in stellar mass and color are presented in \S\ref{subsec:colormorph}.

\section{Galaxy-Halo Model}
\label{sec:model}

\subsection{Abundance Matching}
\label{subsec:sham}

We map $\mstar$ onto subhalos using deconvolution abundance matching (SHAM). See Section 3.3 of \citet{behroozi_etal10}, or the appendix of \citet{kravtsov_etal14}, for an account of the mathematics underlying this technique in the presence of scatter. To perform the abundance matching, we use the publicly available code\footnote{\url{https://bitbucket.org/yymao/abundancematching}} developed by Yao-Yuan Mao \citep{lehmann_etal15} that provides a python wrapper around the C kernel developed in \citet{behroozi_etal10}. For the subhalo property used in SHAM, we use $\mpeak,$ the largest value of $\mvir$ ever attained by the subhalo throughout the entire history of the main progenitor halo.  Using the stellar mass function provided in \citet{moustakas_etal13}, we model $\mstar$ as a log-normal distribution with $0.2$ dex of scatter about the median relation $\median{\mstar}{\mpeak}$ determined by SHAM \citep{reddick_etal13}. As described in the appendix, we include a prescription for supplementing the ordinary {\tt Rockstar} subhalo catalog with disrupted subhalos, so that some of our model galaxies inhabit subhalos that are no longer resolved and do not appear in the standard publicly available catalog (``orphan galaxies").

\begin{figure}
\centering
\includegraphics[width=8cm]{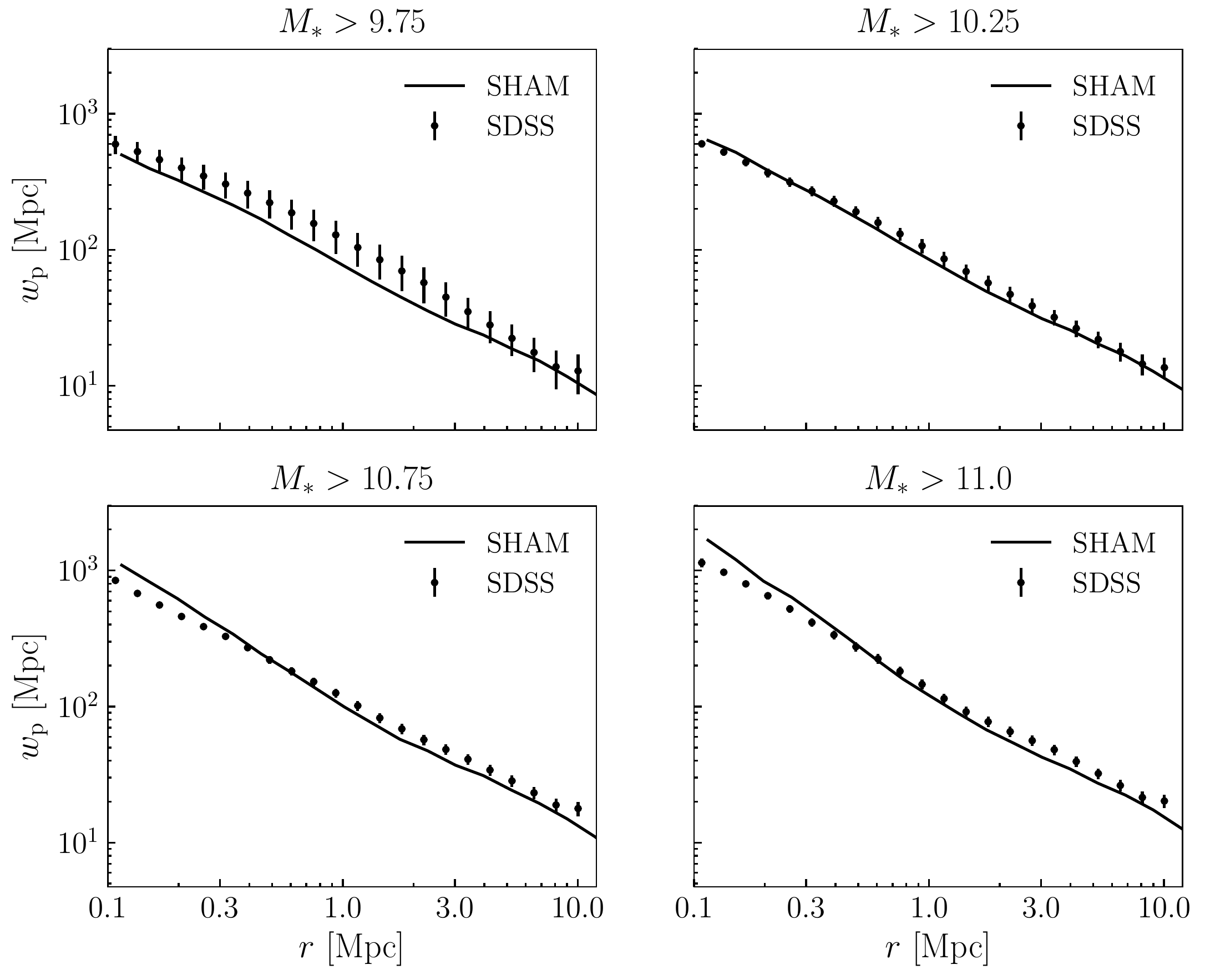}
\caption{
{\bf Abundance matching clustering predictions.}
We show the $\mstar-$dependence of projected galaxy clustering, $\wproj(\rproj),$ comparing standard abundance matching to SDSS galaxies (see \S\ref{subsec:sham} for details). Each panel shows the comparison for a volume-limited sample defined by a different $\mstar-$threshold. Black points with error bars show SDSS measurements; solid black curves show the abundance matching prediction of our fiducial model, which includes the effect of orphan subhalos (see Appendix A). Including orphans mitigates the discrepancy in the clustering predicted by traditional, $\mpeak-$based SHAM, though mild tension remains on small scales for $\mstar\gtrsim10^{10.75}\msun,$ and on all scales for $\mstar\gtrsim10^{9.75}\msun.$
}
\label{fig:baseline_sham_clustering}
\end{figure}

Figure \ref{fig:baseline_sham_clustering} shows that our implementation of SHAM produces model galaxies whose projected clustering is in reasonably good agreement with SDSS galaxies. We note, however, that small-scale clustering shows mild $\sim10-20\%$ tension at stellar mass $\mstar\gtrsim10^{10.75}\msun,$ and similarly on all scales for the $\mstar\gtrsim10^{9.75}\msun$ sample, although the statistical significance of this discrepancy is likely rather mild due to the sample variance of SDSS galaxies with $z\lesssim0.05.$

\subsection{Galaxy size models}
\label{subsec:model}

Our primary results in \S\ref{sec:results} show predictions for the $\rhalf-$dependence of galaxy clustering for two classes of empirical models, described in turn in the two subsections below.

\subsubsection{$\mstar$-based model with satellite mass loss}
\label{subsubsec:strippingmodel}

In the first model we assume that stellar mass $\mstar$ controls $\rhalf.$ To implement this model, for simplicity we tabulate $\median{\rhalf}{\mstar}$ directly from the data, rather than pursue a parametric form, so that this model reproduces the observed $\rhalf-\mstar$ relation by construction. In order to ensure that the model has a level of scatter in $\rhalf$ at fixed $\mstar$ that is comparable to the level reported in \citet{somerville_etal17}, our model sizes are drawn from a log-normal distribution with $0.2$ dex of scatter centered at the observed $\median{\rhalf}{\mstar}.$ 

It is natural to consider the possibility that stellar mass is stripped from satellite galaxies after infall. The basis for our implementation of this phenomenon is the fitting function presented in \citet{smith_etal16}, which was calibrated by studying stellar mass loss in a suite of high-resolution hydrodynamical simulations. In this model, $f_{\ast}$ quantifies the remaining fraction of stellar mass as a function of $f_{\rm DM},$ the fraction of dark matter mass that remains after subhalo infall:
\beq
f_{\ast} = 1 - \exp(-14.2f_{\rm DM}).
\eeq
For $f_{\rm DM}$ we use the ratio of present-day subhalo mass divided by the peak mass, $M_{\rm vir}/M_{\rm peak}.$ If we denote the post-stripping stellar mass as $M_{\ast}',$ then we have $M_{\ast}'\equiv f_{\ast}M_{\ast},$ where $M_{\ast}$ is given by the initial application of abundance matching. We then calculate the post-stripping radius by interpolating $\langle\rhalf'\vert\mstar'\rangle$ directly from SDSS data. In this way, we model satellite mass loss and  size truncation in a manner that mimics what is seen in hydrodynamical simulations. As shown in \citet{smith_etal16}, stellar stripping in this model does not begin until $f_{\rm DM}\lesssim0.2,$ and in order for $\sim50\%$ of stars to be stripped, the DM halo needs to be stripped to $\sim5\%$ of its initial mass. 

\subsubsection{$\rvir-$based model}
\label{subsubsec:rvirmodel}

\begin{figure}
\centering
\includegraphics[width=8cm]{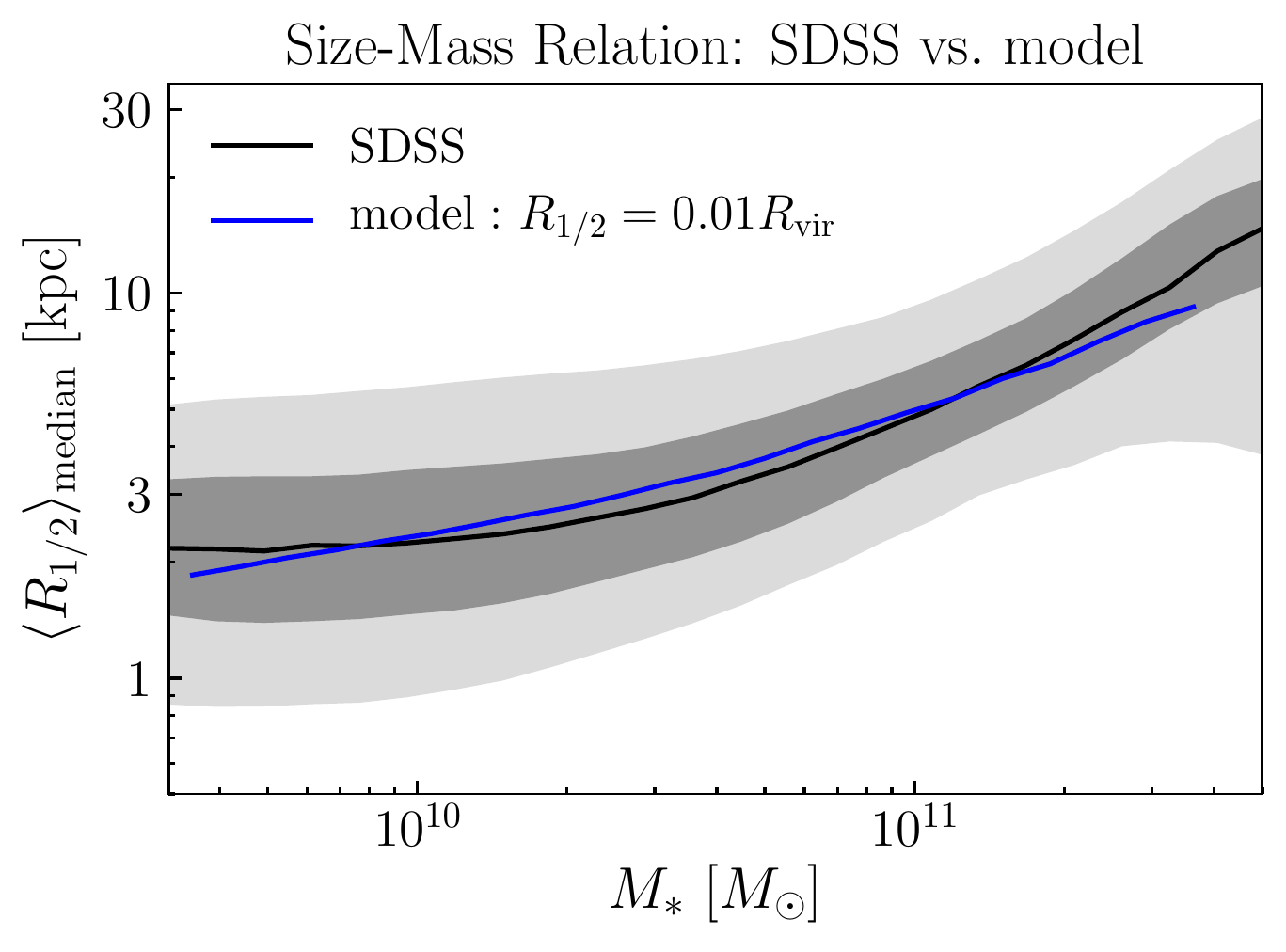}
\caption{
The black curve shows the median size-mass relation of SDSS galaxies as measured in \citet{meert_etal15}. The two gray bands enveloping the black curve show the $50\%$ and $90\%$ percentile regions. The blue curve shows the $\rvir-$based model in which $\median{\rhalf}{\rvir}=0.01\rvir$ (see Section \ref{subsubsec:rvirmodel} for details).  
}
\label{fig:scatter_plot}
\end{figure}

As noted in \S\ref{sec:intro}, it was shown in \citet{kravtsov13} that galaxy sizes are on average proportional to the virial radius of their host halo. Motivated by these results, we  explore a model in which $\rhalf$ scales linearly with halo virial radius:\footnote{We note that the normalization factor in \citet{kravtsov13} is 0.015, whereas our normalization is 0.01. The difference derives from two factors: {\em i.} the choice of halo radius definition: \citet{bryan_norman98} in the present work, compared to the $200c$ definition used in \citet{kravtsov13}; {\em ii.} the latter study used deprojected (larger) sizes for spheroidal galaxies, whereas we make no such correction here. See \S\ref{sec:future} for further discussion.}
\beq
\label{eq:fiducial_model}
\rhalf = 0.01\rvir
\eeq
For the virial radius of halos and subhalos, we use $\rmpeak:$ the value of $\rvir$ in physical units of $\kpc$ measured at the time when halo mass reached its maximum, $\zpeak.$  The relationship between $\rmpeak, \mpeak$ and $\zpeak$ is given by
\beq
\mpeak\equiv\frac{4\pi}{3}\rmpeak^{3}\Delta_{\rm vir}(\zpeak)\rho_{\rm m}(\zpeak),
\eeq
where for $\Delta_{\rm vir}(\zpeak)$ we use the fitting function to the ``virial" definition used in \citet{bryan_norman98}. This modeling assumption effectively defines $\median{\rhalf}{\mhalo};$ this relation defines the center of the log-normal distribution with $0.2$ dex of scatter that we use to generate a Monto Carlo realization of the model population.

\begin{figure*}
\centering
\includegraphics[width=11cm]{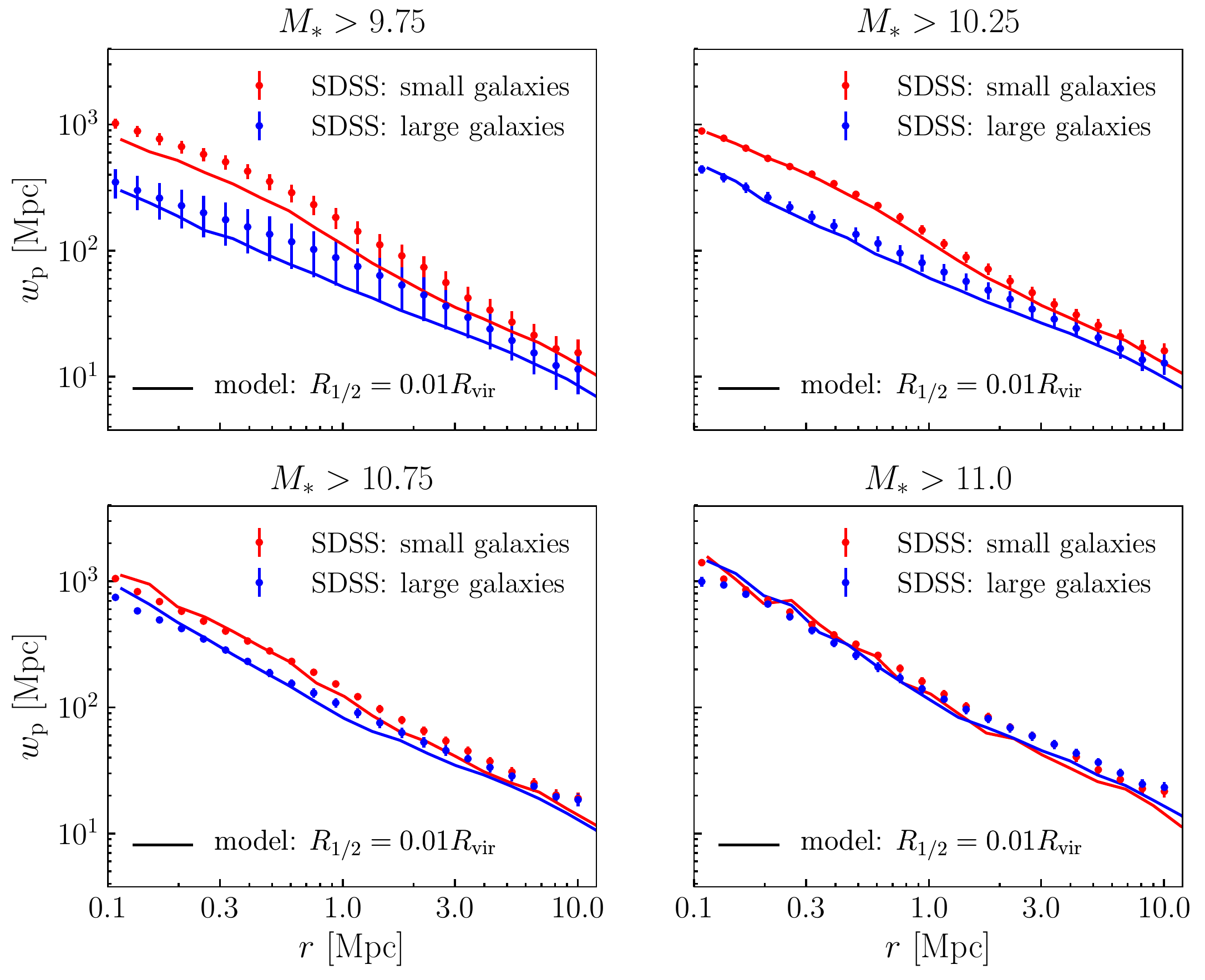}
\caption{
{\bf $\rhalf-$dependence of galaxy clustering: comparison to $\rvir$-based model.}
Red and blue points with error bars show our SDSS measurements of the clustering of small and large galaxies, respectively. For each volume-limited sample of $\mstar-$complete galaxies, the small and large subsamples have identical stellar mass functions, as shown in Figure \ref{fig:sizedefinition}. Small galaxies cluster much more strongly relative to large galaxies of the same stellar mass. Solid curves show the clustering predictions of the $\rvir-$based model described in \S\ref{subsubsec:rvirmodel}. The $\rvir-$based model inherits the shortcoming of ordinary abundance matching shown in Figure \ref{fig:baseline_sham_clustering}, although the model faithfully captures the {\em relative} clustering of small vs. large galaxies, as shown in Figure \ref{fig:clustering_ratio_upshot}.
}
\label{fig:rvir_only_clustering_absolute}
\end{figure*}

\begin{figure*}
\centering
\includegraphics[width=11cm]{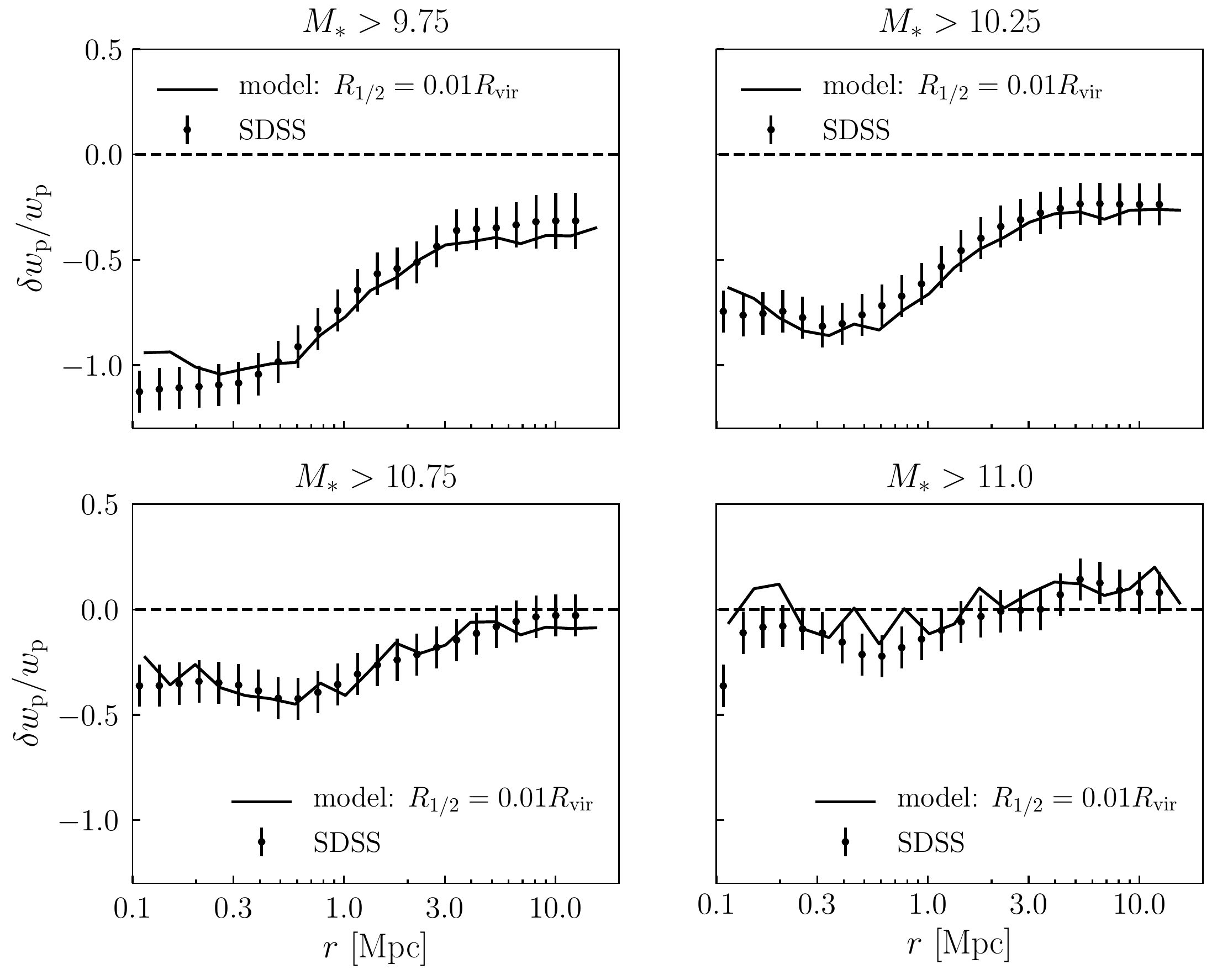}
\caption{
{\bf $\rhalf-$dependence of galaxy clustering: comparison to $\rvir$-based model.}
Closely related to Figure \ref{fig:rvir_only_clustering_absolute}, y-axes show {\em clustering strength ratios,} defined as $\delta\wproj/\wproj\equiv(w_{\rm p}^{\rm large} - w_{\rm p}^{\rm small})/w_{\rm p}^{\rm all}.$ Thus a y-axis value of $-0.5$ corresponds to small galaxies being $50\%$ more strongly clustered than large galaxies of the same stellar mass. Solid curves show predictions of the $\rvir-$based model described in \S\ref{subsubsec:rvirmodel}. Normalizing the measurements and predictions by $w_{\rm p}^{\rm all}$ scales away the shortcoming of ordinary abundance matching (see Figure \ref{fig:baseline_sham_clustering}), highlighting the successful prediction of the $\rvir-$based model for the $\rhalf-$dependence of galaxy clustering.
}
\label{fig:clustering_ratio_upshot}
\end{figure*}

\section{Results}
\label{sec:results}

\subsection{Size-Mass Scaling Relation}
\label{subsec:one_point_function}

In Figure \ref{fig:scatter_plot} we show the scaling of galaxy size $\rhalf$ with $\mstar.$ The black curve enveloped by the gray bands shows the scaling relation for SDSS galaxies, while the blue curve shows the median relation  $\median{\rhalf}{\mstar}$ predicted using the $\rvir-$based model described in \S\ref{subsubsec:rvirmodel}. While the curvature of the median relation in the model is slightly weaker than SDSS, this figure shows that models in which $\rhalf\propto\rvir$ can nonetheless match the observed $\rhalf-\mstar$ relation remarkably well given only a single parameter, confirming the results of \citet{kravtsov13} in a forward modeling context.

\subsection{Size-Dependent Clustering Measurements}
\label{subsec:clustering_results}

In Figure~\ref{fig:rvir_only_clustering_absolute}, we present new measurements of the $\rhalf-$dependence of projected galaxy clustering, $\wproj(\rproj).$ We measure $\wproj(\rproj)$ separately for large and small subsamples for four different $\mstar$ thresholds: an  $\mstar>10^{9.75}\msun$ sample with $z_{\rm max}=0.0435$ and $N_{\rm gal}=27658;$ an  $\mstar>10^{10.25}\msun$ sample with $z_{\rm max}=0.0663$ and $N_{\rm gal}=48571;$ 
an  $\mstar>10^{10.75}\msun$ sample with $z_{\rm max}=0.1$ and $N_{\rm gal}=52350;$ 
and an  $\mstar>10^{11}\msun$ sample with $z_{\rm max}=0.1$ and $N_{\rm gal}=33578.$ For each threshold, we split the galaxies into ``large" and ``small" subsamples, as described in \S\ref{subsec:sizedef}. As shown in Figure \ref{fig:sizedefinition}, our size-based samples  have similar stellar mass functions. Red points with jackknife-estimated error bars show SDSS measurements of $\wproj(\rproj)$ for samples of small galaxies, while blue points show the measured correlation functions for samples of large galaxies. Solid curves show $\wproj$ as predicted for these subsamples by the $\rvir-$based model of galaxy sizes described in \S\ref{subsubsec:rvirmodel}.

The salient feature of these clustering measurements is that small galaxies cluster more strongly than large galaxies of the same stellar mass in all but the largest $\mstar$ samples. Remarkably, both the trend and the quantitative difference in the shape of the correlation functions of small and large galaxy samples are reproduced by the $\rvir-$based model. This may be surprising, since $\rhalf\propto\rvir,$ halo mass $\rvir\propto\mhalo^{1/3},$ and clustering strength of halos increases with $\mvir.$ Based on this simple argument, one would expect that large galaxies would be the more strongly clustered. We provide a resolution to this conundrum in \S\ref{subsec:censat_sizes}. However,  before doing so, we first examine the clustering test of the $\rvir-$based model in more detail.

As shown in Figure \ref{fig:baseline_sham_clustering}, the abundance matching prediction for $\wproj(\rproj)$ exhibits tension with SDSS observations at the $10-20\%$ level. This tension is inherited by our $\rvir-$based model for size, which is the subject of this work, and so we wish to compare our size models to data in such a way that minimizes the role played by the underlying stellar-to-halo-mass relation. We accomplish this using the {\em $\rhalf$ clustering ratios,} described below.

For each volume-limited $\mstar-$threshold sample, we additionally measure $\wproj(\rproj)$ {\em without} splitting on size, giving us measurements $\wpall, \wplarge,$ and $\wpsmall$ for each threshold sample. This allows us to compute the ratio $(\wplarge-\wpsmall)/\wpall,$ which we refer to as {\em the $\rhalf$ clustering ratio}. These ratios, denoted as $\delta\wproj/\wproj,$ are the measurements appearing on the y-axis in each panel of Figure \ref{fig:clustering_ratio_upshot}. As a specific example, a clustering ratio of $-0.5$ corresponds to small galaxies being $50\%$ more strongly clustered than large galaxies of the same stellar mass.

The points and curves in Figure \ref{fig:clustering_ratio_upshot} are almost all negative: small galaxies cluster more strongly relative to large. This presentation of the measurement makes it clear that the underlying signal of $\rhalf-$dependent clustering is strongest for samples with smaller stellar mass; as $\mstar$ increases, the signal weakens and nearly vanishes for $\mstar\gtrsim10^{11}\msun.$ Strikingly, the $\rvir-$based model reproduces this $\mstar-$dependent behavior, as well as the scale-dependence of the observed clustering signal at each $\mstar.$ As described in \S\ref{subsec:censat_sizes} below, we attribute the success of this prediction to the systematic difference between the sizes of central and satellite galaxies.

\subsection{Testing the Impact of Satellite Mass Loss}
\label{subsec:mstar_stripping}
\begin{figure*}
\centering
\includegraphics[width=11cm]{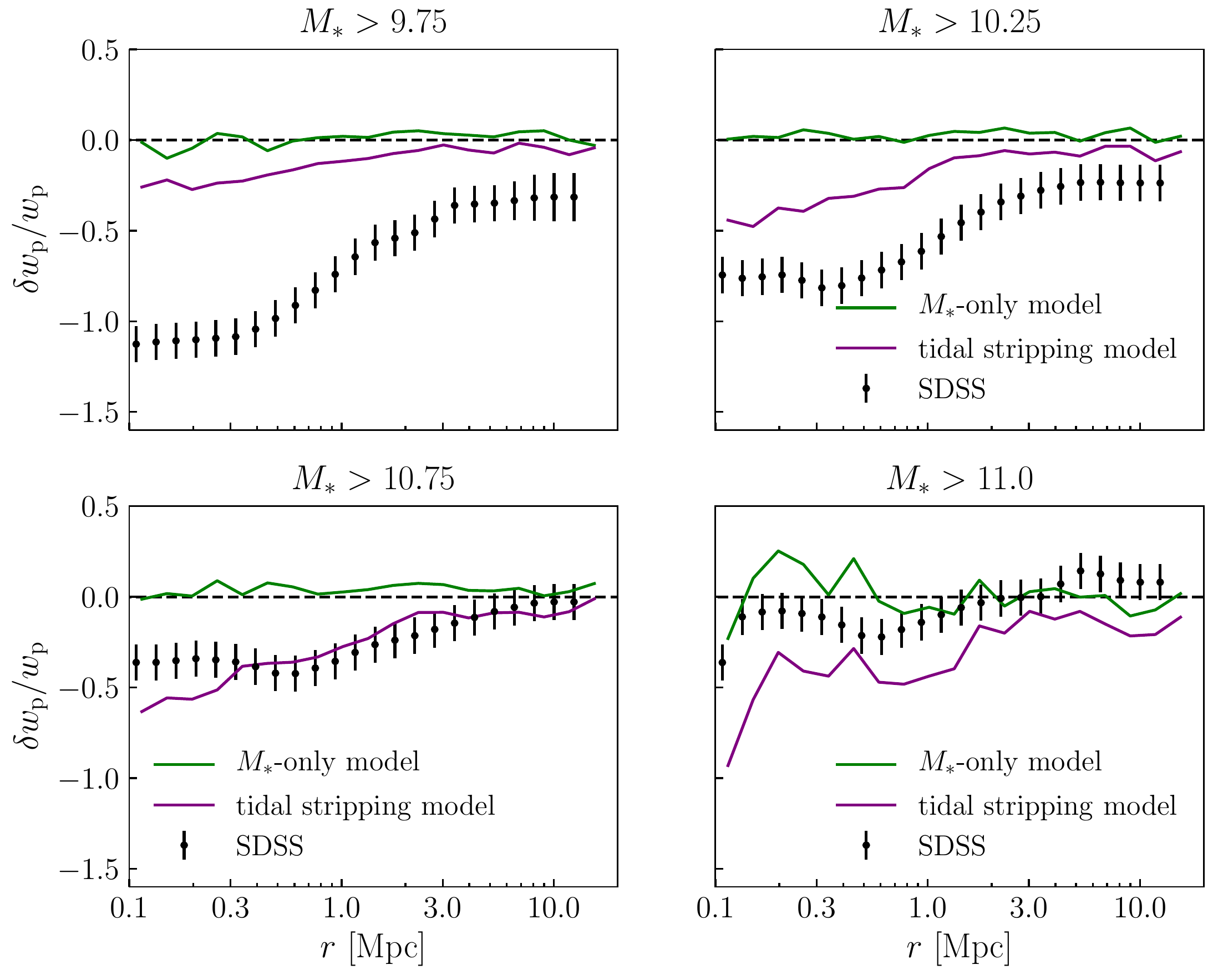}
\caption{
{\bf The subdominant role of tidal stripping.}
In all panels, the axes and points with error bars are the same as in Figure \ref{fig:clustering_ratio_upshot}. The solid green curves show the prediction of a model where $\rhalf$ is statistically set by $\mstar,$ in the complete absence of satellite mass loss. Such a model predicts zero $\rhalf-$dependence to galaxy clustering, in gross tension with observations. The solid purple curves show results for a model in which satellites lose mass after infall in a manner similar to what is seen in high-resolution hydrodynamical simulations, as described in \S\ref{subsubsec:strippingmodel}. This produces satellites that are smaller than centrals, but the effect is too mild to correctly capture the observed clustering. Evidently, satellite-specific mass stripping plays a sub-dominant role in setting the relative size of centrals and satellites.
}
\label{fig:mstarmodelclustering}
\end{figure*}

Figure~\ref{fig:mstarmodelclustering} is analogous to Figure \ref{fig:clustering_ratio_upshot}, but instead compares observed clustering ratios to those predicted by models
in which galaxy size is controlled by $\mstar$ (\S\ref{subsubsec:strippingmodel}). Green solid lines show predictions of the $\mstar$-based model that does not account for mass loss due to tides. In this model, $\rhalf$ follows a purely random log-normal distribution, with the median relation $\median{\rhalf}{\mstar}$ directly matched to the observed relation by construction.  This model thus predicts no dependence of galaxy clustering upon $\rhalf$ at fixed $\mstar,$ simply because the scatter of $\rhalf$ about $\median{\rhalf}{\mstar}$ is uncorrelated with any other variable. The figure shows that this model clearly is at odds with the significant dependence of clustering on size exhibited by SDSS galaxies.

Including satellite mass loss in the model introduces correlations in the scatter: after mass loss, satellites have smaller sizes than centrals of the same $\mstar$ due to post-infall tidal stripping.
The purple curves in Figure \ref{fig:mstarmodelclustering} show that satellite mass loss impacts $\rhalf-$dependent clustering in the expected manner. The ``small" galaxy population has a higher satellite fraction due to satellites being smaller than centrals of the same $\mstar,$ and so in this model small galaxies cluster more strongly relative to large. However, the magnitude of the effect is not sufficiently strong  to produce clustering predictions that are consistent with observations. 

Due to the shape of ${\rm d}\rhalf/{\rm d}\mstar,$ low-mass satellites have to lose much more halo mass than high-mass satellites in order to get pushed into the small-$\rhalf$ sample. Thus even though there are more satellites at lower masses, the shape of ${\rm d}\rhalf/{\rm d}\mstar$ counteracts this trend, making the mass-dependence of the observed signal shown in Figure \ref{fig:mstarmodelclustering} difficult to explain by appealing to tidal stripping. Evidently, it is difficult to strip enough mass from satellites so that the $\rhalf-$dependent clustering predictions are in agreement with observations.  We reach the same conclusion when we tested an even more extreme $\mstar-$stripping model, in which stellar mass loss is linearly proportional to halo mass loss \citep[][Model 1]{watson_etal12}.
As we argue in \S\ref{sec:interpretation}, this supports the conclusion that the relative difference between median sizes of central and satellite galaxies is largely in place at the time of satellite infall.

\subsection{The Role of Galaxy Color}
\label{subsec:colormorph}

We now examine how galaxy clustering depends on size when we control both for stellar mass and color dependence of galaxy clustering. Using a volume-limited galaxy sample with $10^{9.75}\msun < \mstar< 10^{10.25}\msun,$ we first divide the sample into ``red" and ``blue" subsamples using $g-r=0.65$ as a boundary, which approximately corresponds to the trough of the green valley for this stellar mass. For each color-selected subsample, we separately measure $\median{\rhalf}{\mstar; {\rm red}}$ and $\median{\rhalf}{\mstar; {\rm blue}},$ and use these median size relations to split each subsample into two.

\begin{figure}
\centering
\includegraphics[width=8cm]{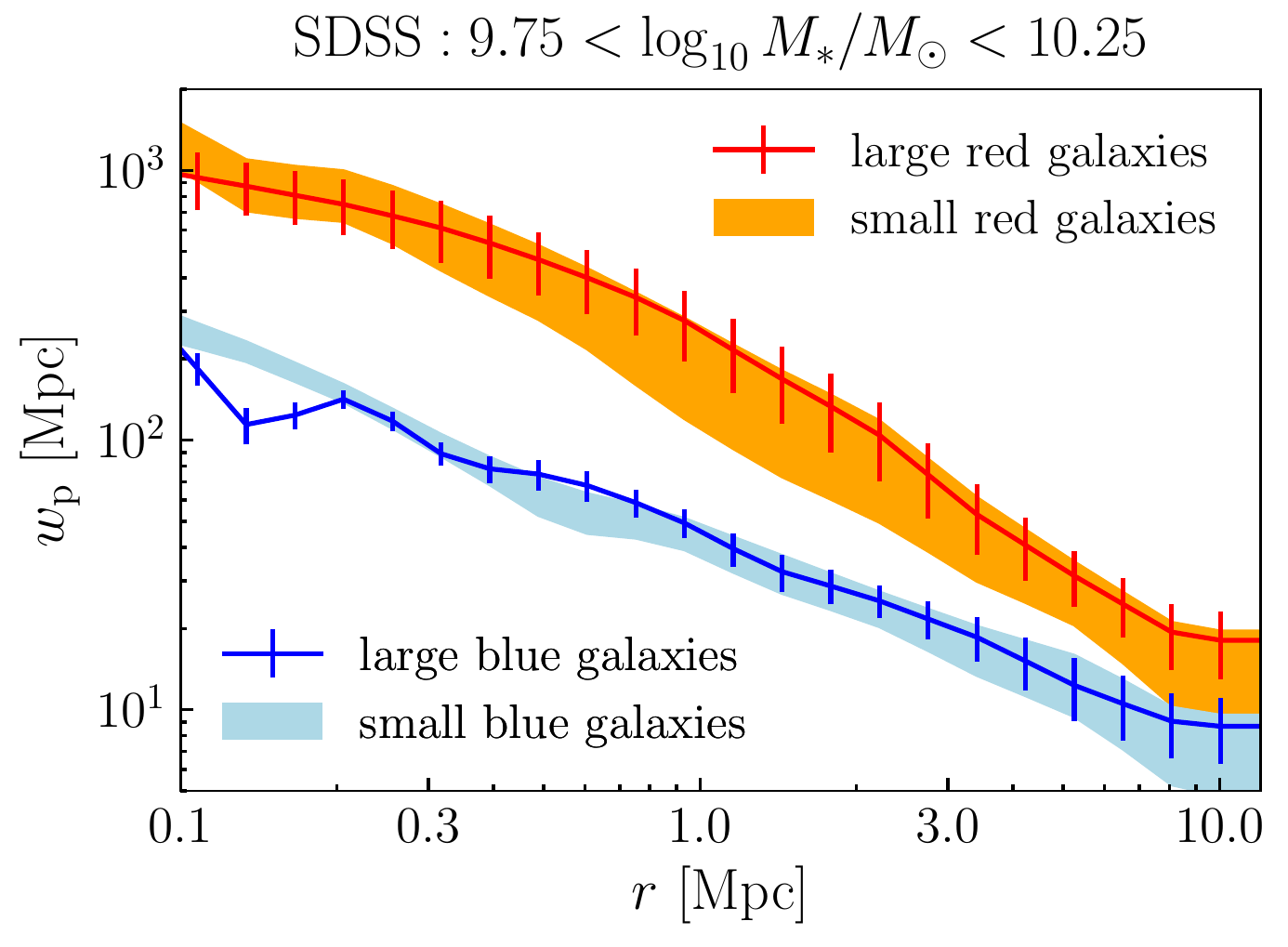}
\caption{
{\bf Distinct $\rhalf-$dependence of clustering for color-selected galaxy samples.}
Projected clustering of SDSS galaxies in the bin of stellar mass: $10^{9.75}<\mstar/\msun<10^{10.25}.$ For both ``blue" and ``red" samples, we first divide SDSS galaxies falling in the $\mstar$ bin based on a color cut at $g-r=0.65.$ We then subsequently compute $\median{\rhalf}{\mstar; {\rm red}}$ and $\median{\rhalf}{\mstar; {\rm blue}}$. In analogy to the bottom left panel of Figure \ref{fig:sizedefinition}, for each of the red and blue samples, the small and large subsamples have statistically indistinguishable stellar mass functions. We then measure $\wproj$ for each of the four subsamples; jackknife-estimated errors are shown alternately using error bars and shaded bands, as specified in the legend. At this stellar mass, pre-selecting galaxy samples by color results in a dramatic weakening of the $\rhalf-$dependence of $\wproj$ relative to $\mstar-$complete samples, as well as a sign-reversal of the signal. In \S\ref{sec:interpretation} we give a simple halo model picture unifying the trends shown here with those illustrated in Figure \ref{fig:clustering_ratio_upshot}.
}
\label{fig:colorclustering}
\end{figure}

Figure \ref{fig:colorclustering} shows the clustering of large vs. small galaxies, separately for red and blue samples. For each color-selected sample, large galaxies are slightly more strongly clustered relative to small galaxies. Comparing Figure \ref{fig:colorclustering} to the the upper right panel of Figure \ref{fig:rvir_only_clustering_absolute} shows the size dependence for color-selected samples is dramatically weaker.   For $\mstar-$complete samples, small galaxies cluster much stronger than large galaxies; for color-selected samples, the reverse is true, but the difference is much smaller and is confined to a limited range of scales.  Although the two results may seem surprising, even contradictory, they can, in fact be explained within a simple, unified model, as will be discussed in \S\ref{sec:interpretation}.

\begin{figure}
\centering
\includegraphics[width=8cm]{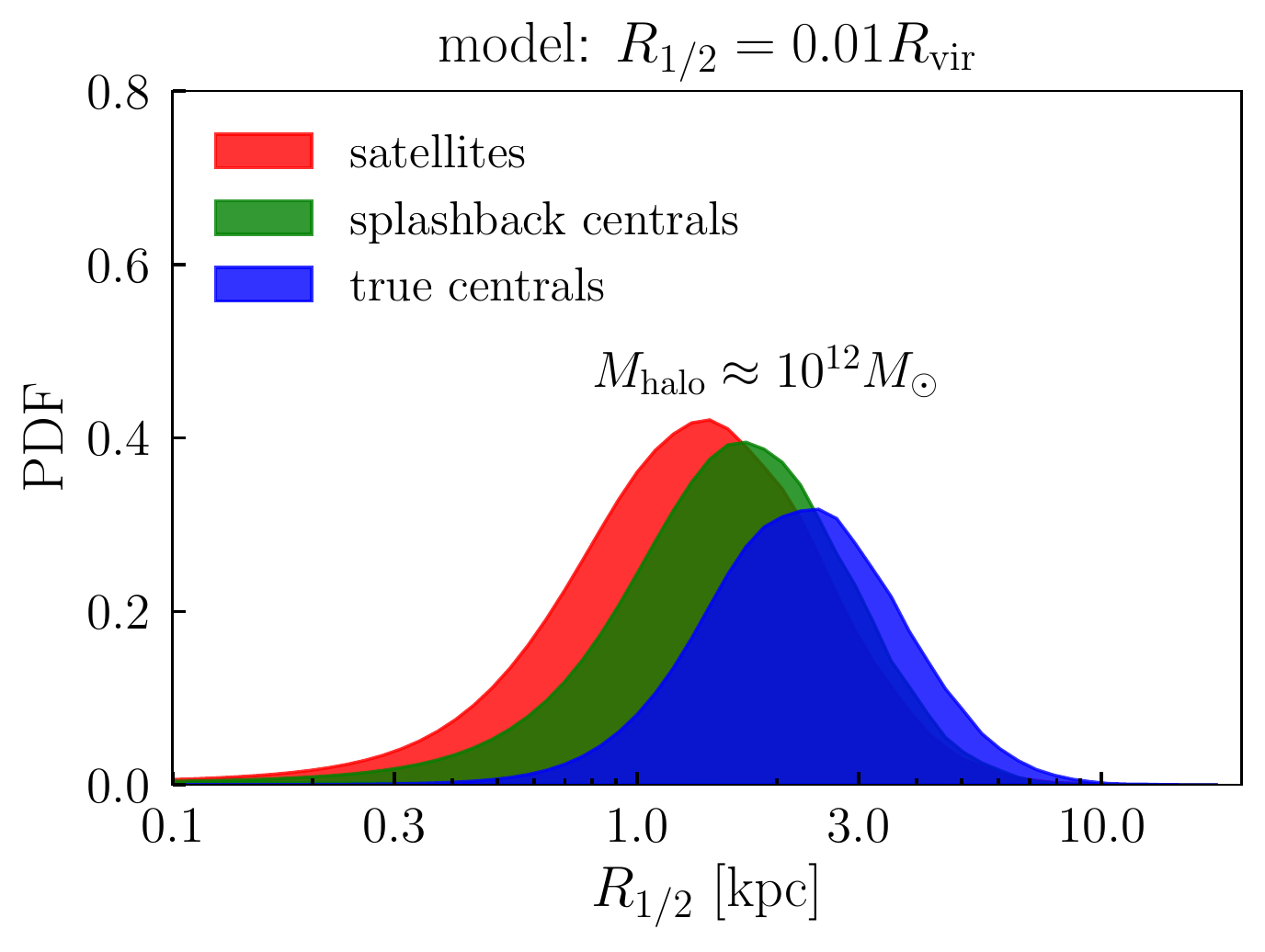}
\caption{
{\bf Relative sizes of centrals and satellites.}
In a narrow bin of subhalo mass $\mhalo\equiv\mpeak\approx10^{12}\msun,$ we show the distribution of galaxy sizes for different subpopulations, as predicted by the $\rvir-$based model. The red distribution shows the sizes of satellites; the blue distribution shows host halos that have never passed inside the virial radius of a larger halo (``true centrals"); the green distribution host halos that were subhalos inside a larger halo at some point in their past history (``splashback centrals"). In the $\rvir-$based model, galaxy size is set by the {\em physical} size of the virial radius at the time the halo attains its peak mass, naturally resulting in smaller sizes for satellites and splashback centrals relative to true centrals of the same mass.
}
\label{fig:censatsizehist}
\end{figure}

\section{A Unified Interpretation of $\rhalf-$Dependent Clustering}
\label{sec:interpretation}

In the previous section, comparing model predictions to the observed size dependence of galaxy clustering produced two intriguing results. First, the model that assumes $\rhalf\propto\rvir$ predicts with remarkable accuracy the dependence of both the amplitude and shape of  size-dependent galaxy correlation functions, especially considering that the model nominally has only one free parameter: the normalization of the linear $\rhalf-\rvir$ relation. At the same time, small galaxies cluster more strongly relative to large, which seems at odds with intuitive expectations that, within such a model, larger sizes correspond to larger halo masses, $\rvir\propto\mvir^{1/3}$; thus  naively, galaxies with larger sizes should be more strongly clustered, not less. Second, the strong dependence of galaxy clustering on size weakens to almost no dependence for samples of red and blue galaxies selected by color within a given stellar mass range.
Below, in \S\ref{subsec:censat_sizes}, we propose a simple and unified explanation of all these results within the $\rvir$-based size model; we discuss the broader implications of our interpretation in \S\ref{subsec:broader_implications}.

\subsection{The Relative Size of Centrals vs. Satellites}
\label{subsec:censat_sizes}

Recall that in the $\rvir$-based size model, galaxy size is proportional to the physical size of halo virial radius at $\zpeak,$ the redshift when the halo attained its maximum mass. Thus we can expect that in this model, satellite galaxies will have smaller sizes than  central galaxies hosted by halos of similar mass.  Indeed, Figure \ref{fig:censatsizehist} shows the $\rhalf$ distributions of central, satellite, and ``splashback'' central galaxies living in subhalos of the same mass, $\mhalo\equiv\mpeak\approx10^{12}\msun.$ A ``splashback central"  is defined as a present-day central galaxy that used to be a satellite, i.e., its main progenitor halo passed inside the virial radius of a larger halo at some point in its past history. On the other hand, we define a ``true central" as a galaxy that has never been a satellite. The figure shows that there is a clear trend of sizes decreasing from the central to splashback to satellite galaxies.

In the $\rvir-$based model, satellites and splashback galaxies are smaller than centrals of the same halo mass due to the physical size of their halo being smaller at earlier times $\zpeak.$ There are two distinct reasons why this feature results in small galaxies being more strongly clustered relative to larger galaxies of the same mass. First and foremost, satellite galaxies statistically occupy higher mass host halos that are more strongly clustered. So in models where satellites are smaller than centrals of the same mass, smaller galaxies will have a higher satellite fraction, boosting the clustering of small relative to large galaxies of the same stellar mass. Second, at fixed mass, halos of $L_\ast$ galaxies that form earlier are more strongly clustered, a phenomenon commonly known as {\em halo assembly bias} \citep{gao_white05,wechsler_etal06}. Splashback halos are typically earlier-forming than true centrals \citep{wang_etal09}, and so models where splashback halos host smaller-than-average galaxies will naturally predict smaller galaxies being the more strongly clustered \citep[see, e.g.,][for a demonstration of the splashback-dependence of halo clustering]{sunayama_etal16}.

We can also understand how the observed $\rhalf-$dependent clustering scales with $\mstar$ (shown by comparing different panels in Figure \ref{fig:clustering_ratio_upshot}) in terms of the different sizes of centrals and satellites. The satellite fraction $F_{\rm sat}(\mstar)$ decreases as $\mstar$ increases \citep[e.g.,][]{guo_etal11,reddick_etal13}. As $F_{\rm sat}$ decreases, there are fewer satellites available to preferentially weight the subsample of smaller galaxies. Thus,  high-$\mstar$ samples consist mostly of central galaxies, and the split by size no longer corresponds to the split into satellite- and central-dominated samples. The size-dependence of clustering correspondingly weakens and nearly disappears for samples with large $\mstar$.

Likewise, the significantly weaker dependence of clustering on size for color-selected samples (Figure \ref{fig:colorclustering}), and its reversed sign, may be explained by the same principles.  It is well known that satellites are redder than centrals of the same stellar mass \citep[e.g.,][]{vdB_etal08}, and that the clustering differences between red and blue galaxies can be understood in terms of red satellites preferentially occupying high-mass halos \citep[e.g.,][]{zehavi_etal11}. Pre-selecting galaxies by color then absorbs much of the potential effect that can be reaped by central vs. satellite differences, and so the relative clustering of centrals becomes more important. In the limit that size-dependent clustering is dominated by centrals, the $\rvir$-based size model predicts that large galaxy samples should be (weakly) more clustered than small ones, because $\rhalf\propto\rvir\propto\mhalo^{1/3}.$ This is the trend seen in the clustering measurements shown in Figure \ref{fig:colorclustering}. 

While we consider such a conjecture plausible, a conclusive confirmation requires proper forward modeling of size and color jointly. Such an effort is well motivated by our results, but is beyond the scope of the present work.

\subsection{Broader Implications}
\label{subsec:broader_implications}

The success of the $\rvir$-based model in predicting the size dependence of galaxy clustering may be explained if the sizes of {\it all} galaxies, red, blue, spheroid, disks, satellites and central, are, on average, proportional to the $\rmpeak,$ the physical size of the virial radius at the time at which $\mhalo$ stops growing. Although correlation is not necessarily causation, such a universal and simple dependence poses a useful challenge to fine-grained galaxy formation models, as it indicates a tight connection between the growth of galaxies and their host halos.

The $\rvir$-based size model assumes as an ansatz that galaxy size remains approximately constant after its halo stops growing; as shown in Figure \ref{fig:clustering_ratio_upshot}, this assumption is consistent with the observed size-dependence of galaxy clustering in SDSS.  The halos of typical satellites reach $\mpeak$ far outside the virial radius of their ultimate host halo \citep{behroozi_etal14}, and so in the $\rvir-$based model the size difference between satellites and centrals is largely in place prior to the time of satellite infall. Halos of central and satellite galaxies of the same $\mstar$ differ in the epoch $\zpeak$ at which their halo stopped growing, and this difference gives a natural explanation for their different median sizes.

This difference also explains the relative trends of $\rhalf-\mstar$ relations for red and blue, or disk and spheroid galaxies.
Both types of galaxies exhibit weak dependence of $\rhalf$ on $\mstar$ at $\mstar\lesssim 5\times 10^{10}\ M_\odot$ and
at these masses $\rhalf$ of disks is a factor of two larger than those of spheroids. At larger masses, the sizes of galaxies of both types show stronger dependence on $\mstar,$ and the difference in median size of disk and spheroid galaxies decreases \citep[e.g.,][]{bernardi_etal14}.

In the $\rvir$-based size model, the weak dependence of $\rhalf$ on $\mstar$ for low-mass galaxies is due to the steep $\mstar-\mvir$ relation. For example, if $\mstar\propto\mvir^\alpha\propto\rvir^{3\alpha}$, then $\rhalf\propto \rvir\propto \mstar^{1/(3\alpha)}$. Given that $\alpha\sim 1-3$ is inferred for the $\mstar-\mvir$ relation at low masses \citep{kravtsov10,moster_etal13,behroozi13_smhm,kravtsov_etal14}, the $\rhalf-\mstar$ relation in this regime is expected to be shallow.
For galaxies of $\mstar\gtrsim 5\times 10^{10}\ M_\odot$, $\alpha\approx 0.3$ \citep[e.g.,][]{kravtsov_etal14}, and thus
the slope of the $\rhalf-\mstar$ relation is approximately linear. These trends of $\rhalf$ with $\mstar$ have a direct relation to the trends of surface brightness $\mu_\star=\mstar/(2\pi\rhalf^2)$ with stellar mass \citep[see, for example, the $\mu_\star-$dependent clustering measurements in][which are qualitatively similar to the measurements reported here]{li_etal06}.

The difference in median sizes of disk and spheroidal galaxies at $\mstar\lesssim 10^{11}\ M_\odot$ can be explained by the sizes of satellite galaxies being systematically smaller than the sizes of central galaxies of the same $\mstar,$ coupled with the fact that the satellite fraction is much larger among spheroidal galaxies compared to disk galaxies. For example, the analysis in \citet{rodriguez_puebla_etal15} shows that in this regime, the satellite fraction of blue galaxies is only $\sim10-20\%$, while
the satellite fraction of red galaxies is $\gtrsim 30-50\%,$ increasing with decreasing stellar mass.
As $\mstar$ increases, the satellite fractions of both blue and red galaxies drop, and at $\mstar\gtrsim 10^{11}\ M_\odot,$ the vast majority of blue and red galaxies are centrals. If these galaxies occupy similar halo masses at the same $\mstar$, then the $\rvir$-based size model predicts that the difference in sizes of red and blue galaxies should disappear at these masses, as is indeed observed. A testable prediction of this picture is that when 3D (de-projected) sizes of only central red and blue galaxies of the same $\mstar$ are compared, the sizes should be much closer than the sizes of the overall red and blue populations.

In any model where galaxy $\rhalf$ co-evolves with halo $\rvir,$ the choice of halo boundary definition necessarily impacts the model prediction. Since the clustering of halos has non-trivial dependence upon halo definition \citep{villarreal_etal17}, then measurements of galaxy clustering may be able to observationally distinguish a particular halo boundary that is most closely connected to galaxy size. The redshift evolution of the relationship between $\mhalo$ and $\rvir$ also varies with halo definition, and so measurements of $\rhalf-$dependent clustering across redshift may provide a further observational handle on the halo radius definition that best predicts galaxy size. 

\begin{figure*}
\centering
\includegraphics[width=8cm]{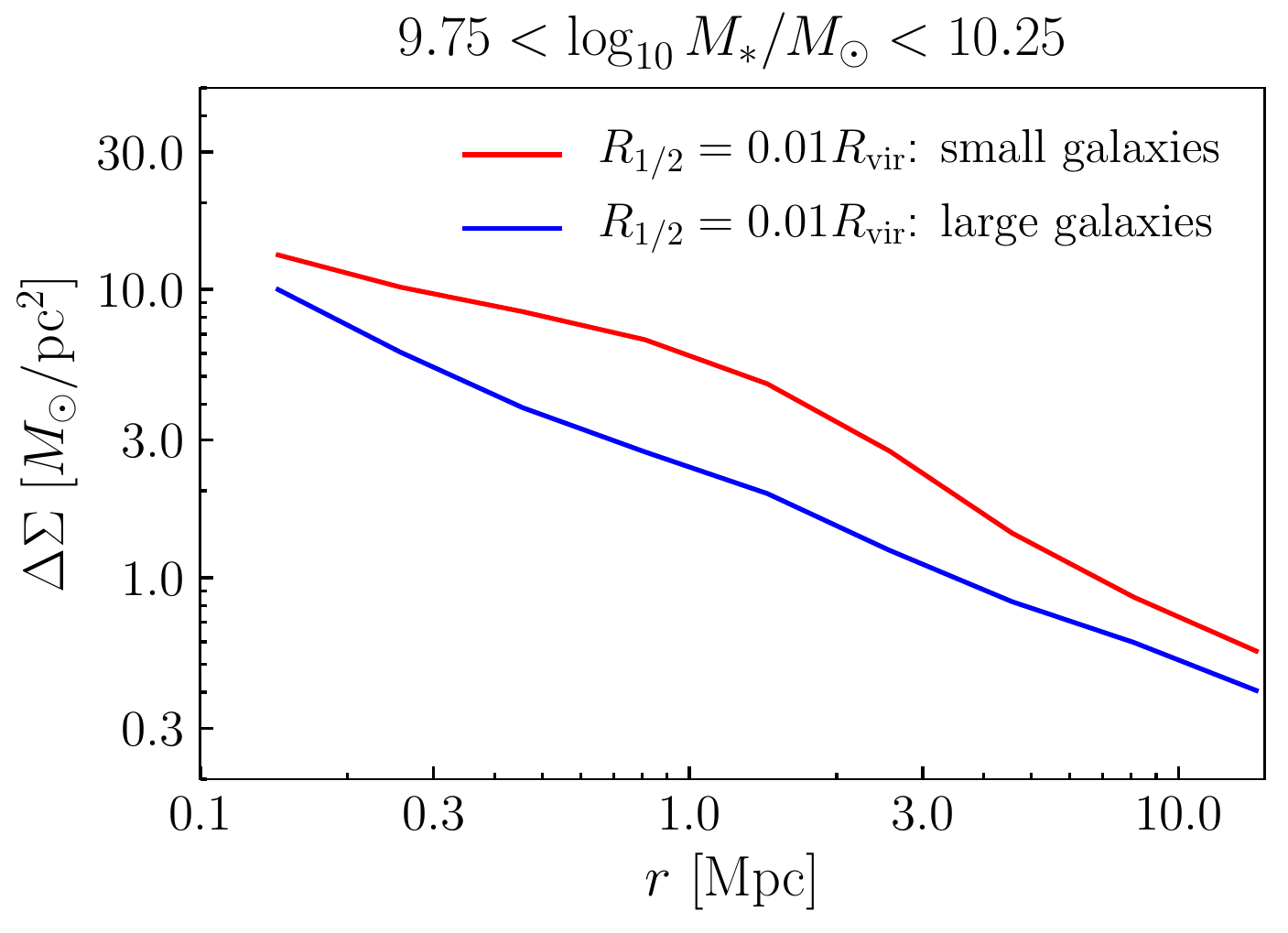}
\includegraphics[width=8cm]{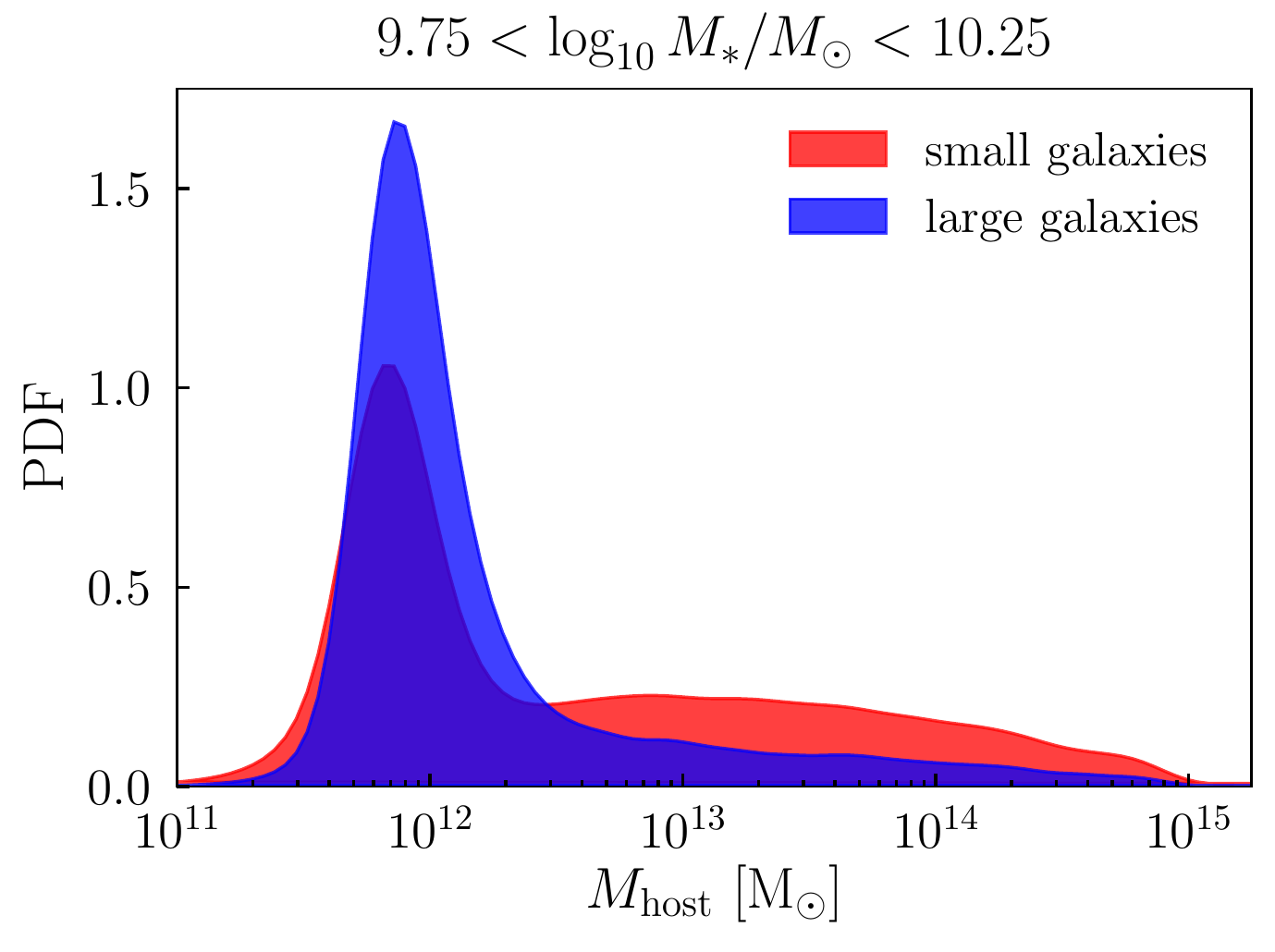}
\includegraphics[width=8cm]{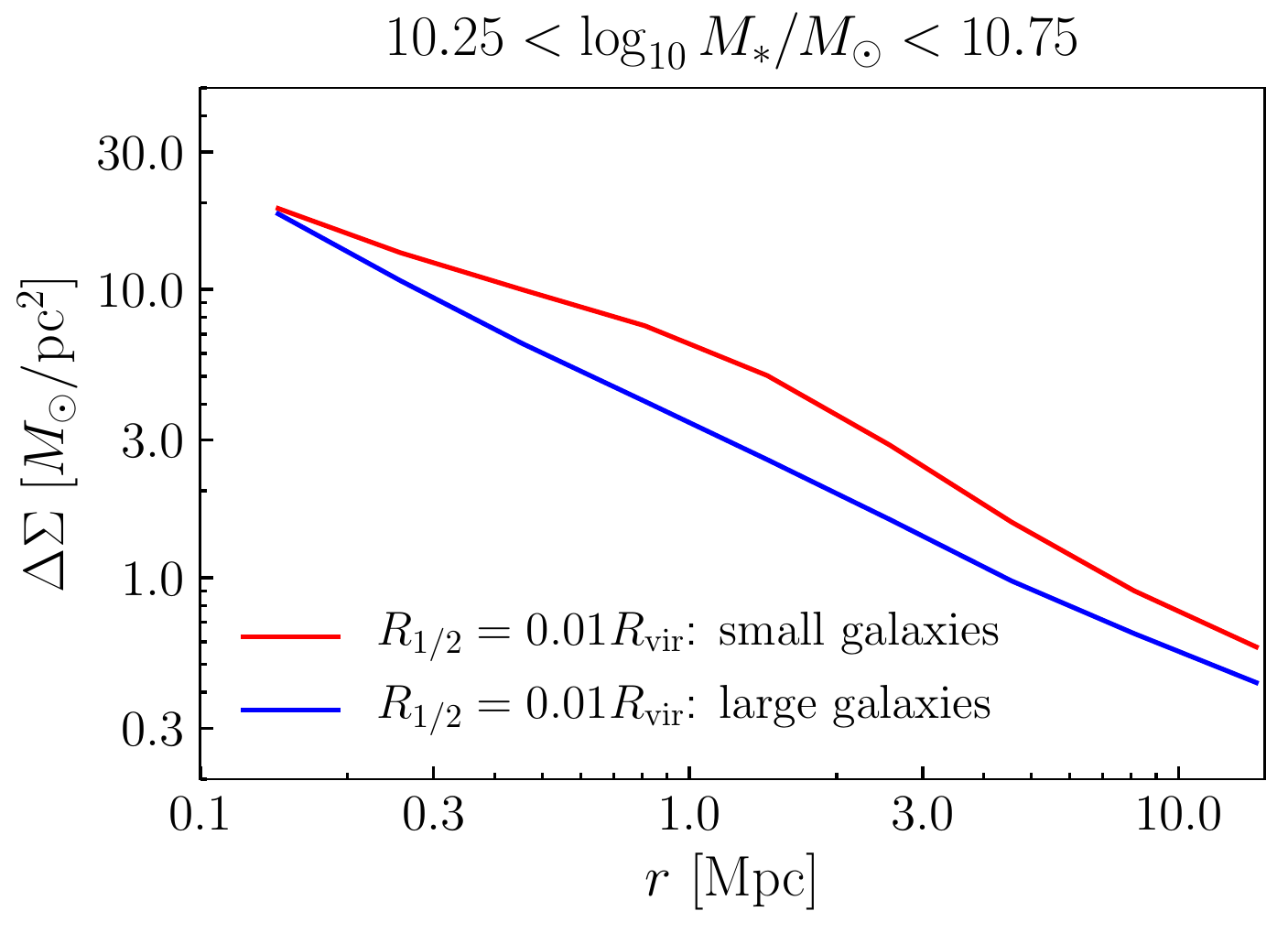}
\includegraphics[width=8cm]{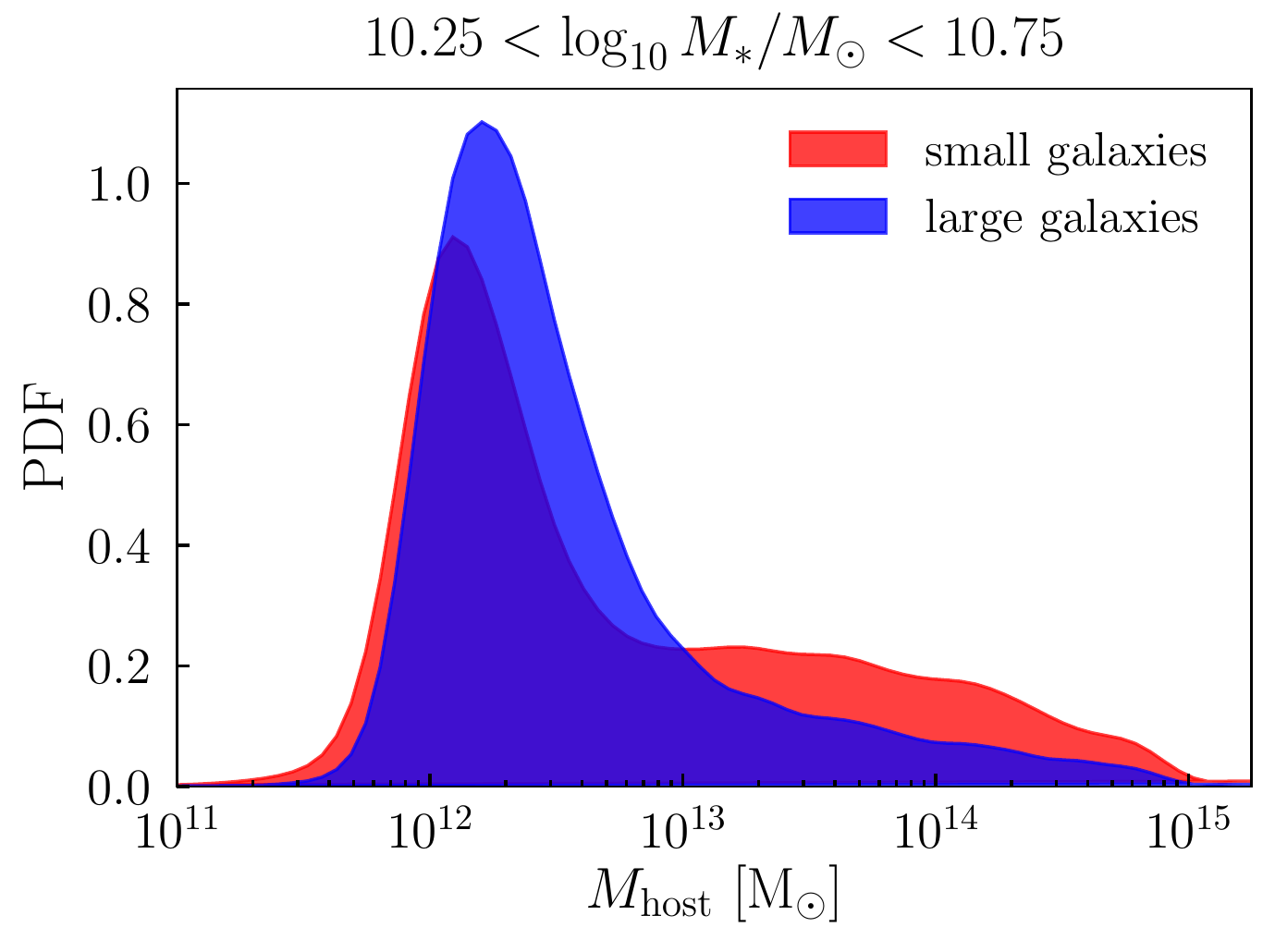}
\includegraphics[width=8cm]{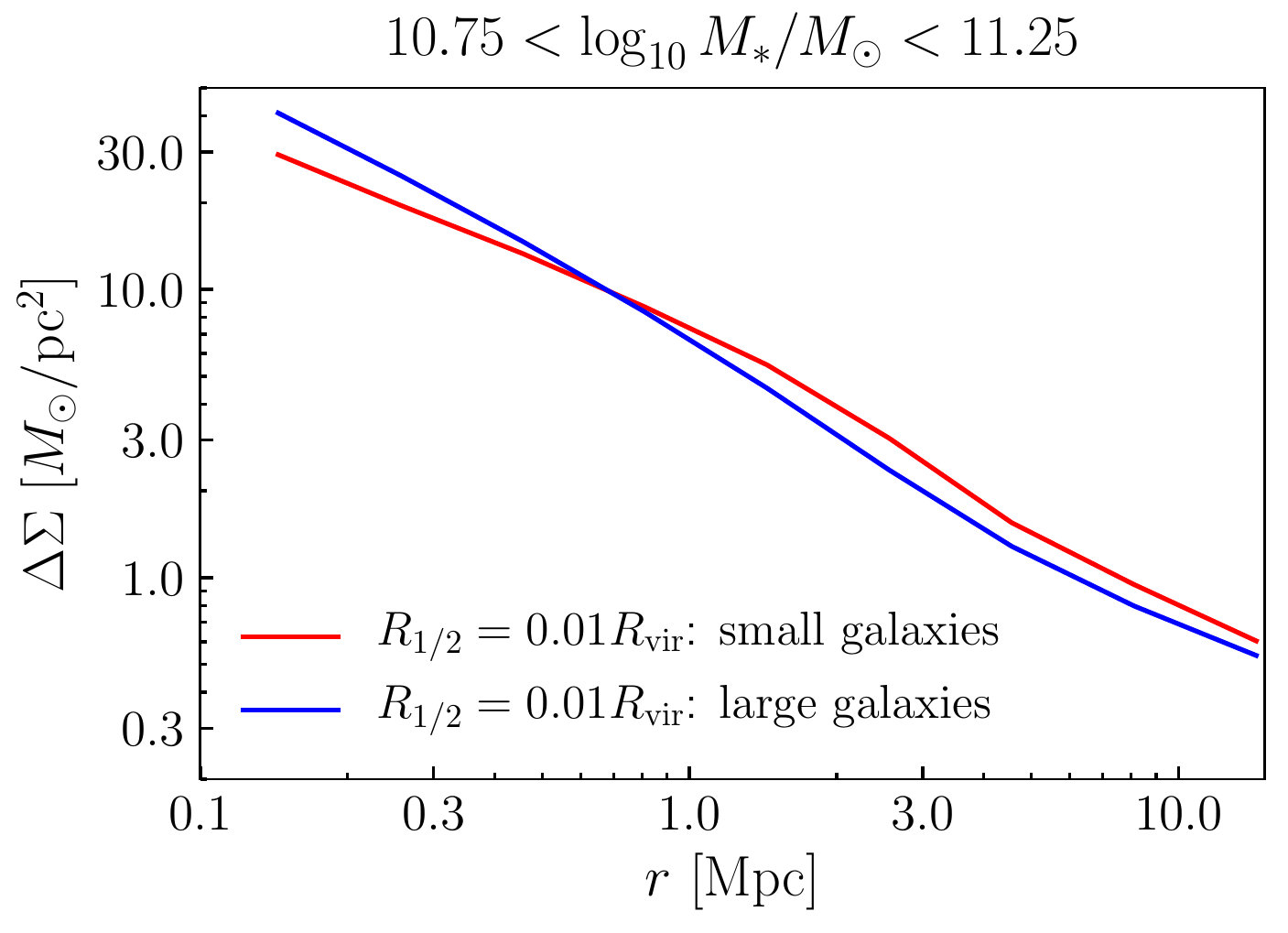}
\includegraphics[width=8cm]{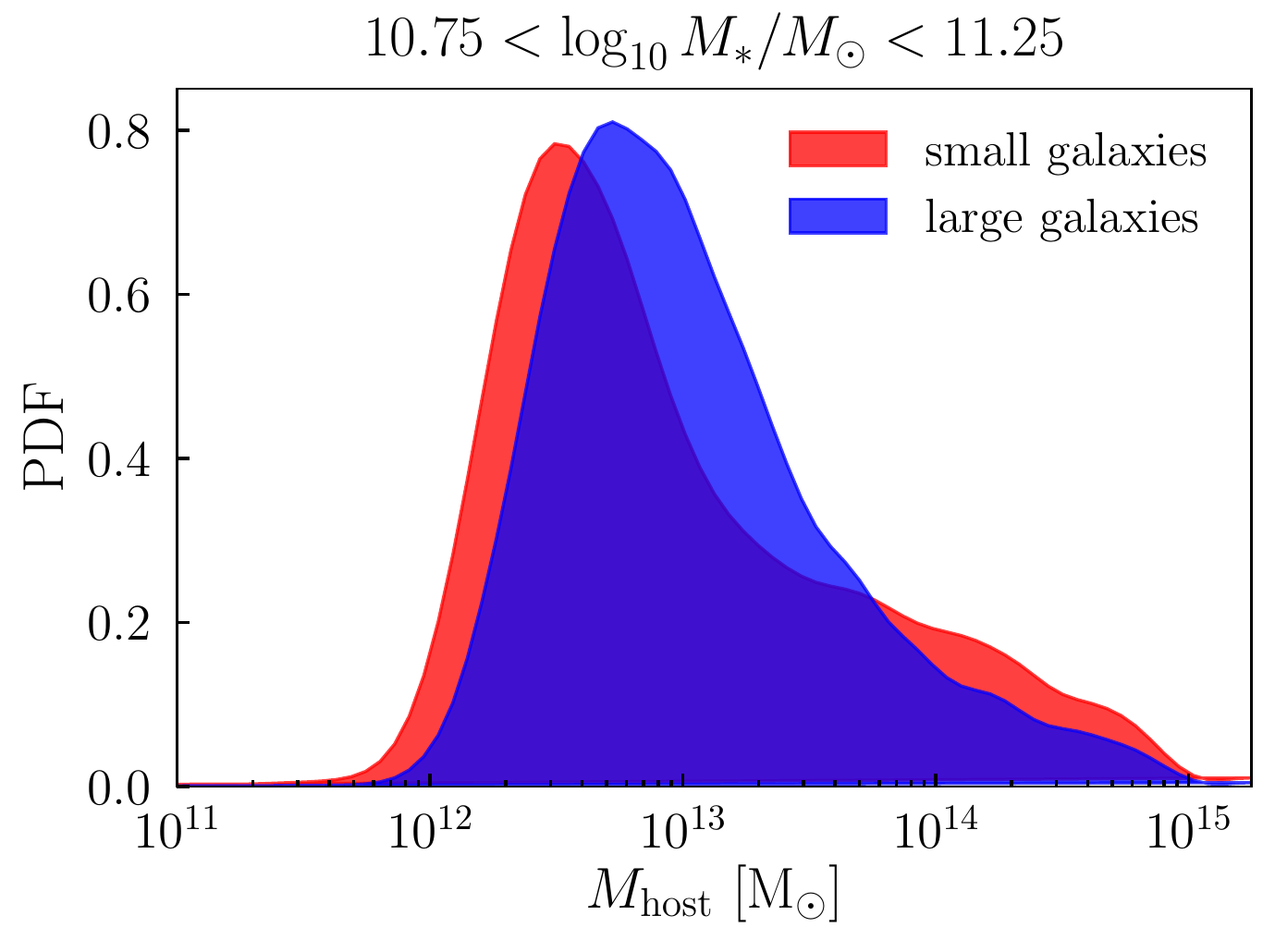}
\caption{
{\bf Prediction for $\rhalf-$dependence of galaxy lensing.}
Using the $\rvir-$based model described in \S\ref{subsubsec:rvirmodel}, in the {\em left panels} we show predictions for future measurements of the $\rhalf-$dependence of galaxy lensing of $\mstar-$complete samples, where ``small" and ``large" subsamples are defined as in \S\ref{subsec:sizedef}. To date, the $\rhalf-$dependence of $\Delta\Sigma$ has only been measured for color-selected samples \citep{charlton_etal17}, in which the observed trends are reversed and weaker than those predicted here: for both blue and red samples in CFHTLenS, the observed $\Delta\Sigma$ of large galaxies is (slightly) stronger relative to small galaxies. As a generic consequence of satellites being smaller than centrals of the same mass, we predict that future lensing measurements of $\mstar-$complete samples will show that $\Delta\Sigma$ of small galaxies is significantly stronger relative to large galaxies of the same stellar mass $\mstar\lesssim10^{11}\msun.$ We additionally predict that at sufficiently large mass $\mstar\gtrsim10^{11}\msun,$ the trend of small-scale $\Delta\Sigma$ with $\rhalf$ reverses sign. The {\em right panels} display the physical explanation for this prediction. For $\mstar\lesssim10^{11}\msun,$ satellites preferentially populating the small sample heavily weight high-mass halos, boosting the lensing signal on all scales. For $\mstar\gtrsim10^{11}\msun,$ when $F_{\rm sat}\lesssim10\%,$ central galaxies dominate the signal, and the $\rvir-$based model predicts that $\rhalf$ increases with $\mhalo$ for centrals. See \S\ref{sec:previous_work} for further details. 
}
\label{fig:lensingprediction}
\end{figure*}

\section{Relation to Previous Work}
\label{sec:previous_work}

Numerous previous analyses have employed the \citet{yang_etal05b} group catalog to study the relationship between galaxy mass, size, and environment. For example, by treating observed groups as genuine dark matter halos, several analyses have found that the $\mstar-\rhalf$ relation of early-type galaxies exhibits weak, if any, environmental dependence \citep{weinmann_etal08,huertas_company_etal13b,shankar_etal14}.

Group-finding methods are subject to significant systematics, the most severe of which are caused by the fracturing of a massive halo's galaxies into multiple small groups, and the fusing of small halos into an apparently large group \citep{duarte_mamon14}. As shown in \citet{campbell_etal15}, these effects can significantly bias attempts to infer trends in the galaxy--halo connection directly from group catalogs. The chief source of these  biases is that the frequency of fracturing and fusing has significant dependence upon large-scale density, as do galaxy properties such as color and morphology. The correlation between group misidentification rate and galaxy properties would be quite challenging to model analytically, but these systematics damage the robustness of the inference: contemporary algorithms {\em generically} wash out differences between centrals and satellites; trends can even be artificially created that are not truly present. 

 Our forward modeling inference technique is not subject to such systematics because our methods are entirely outwith the group-finding framework, and so it is encouraging that our conclusions are commensurable with group-based results. Once galaxies have been color-selected, Figure \ref{fig:colorclustering} shows that the observed clustering trends nearly vanish in magnitude, and reverse sign. Insofar as blue galaxies are largely disk-dominated, our clustering results in Figure \ref{fig:colorclustering} are also consistent with the disk model presented in \citet{dutton_etal08,dutton_etal10}, which predicts that at fixed stellar mass, there is weak, positive correlation between disk size and halo mass. Our conclusions are also qualitatively consistent with \citet{spindler_wake17}, who find that centrals in the \citet{yang_etal05a} group catalog are $\sim20-25\%$ larger than satellites of similar halo mass (using stellar velocity dispersion as an observational proxy for halo mass); we find the same trend at a comparable, if slightly larger magnitude.

The $\rhalf-$dependence of galaxy lensing has recently been measured using CFHTLenS observations \citep{heymans_etal12,erben_etal13}. For both red- and blue-sequence galaxies, it was found in \citet{charlton_etal17} that the lensing signal, $\Delta\Sigma,$ of large galaxies is slightly stronger relative to smaller galaxies. This result is in good agreement with the clustering measurements in Figure \ref{fig:colorclustering}, which show the same trend. 

To date, the $\rhalf-$dependence of $\Delta\Sigma$ has not yet been measured for $\mstar-$complete samples. In the left panels of Figure \ref{fig:lensingprediction}, we show predictions of the $\rvir-$based model for future measurements of $\Delta\Sigma.$ The halo model principles underlying the lensing trends are the same as for $\wproj.$ For $\mstar\lesssim10^{11}\msun,$ the difference in $F_{\rm sat}$ between small and large samples is the dominant factor; satellites occupy higher mass host halos relative to centrals of the same stellar mass, boosting $\Delta\Sigma$ for small relative to large galaxy samples. At sufficiently large stellar mass $\mstar\gtrsim10^{11}\msun,$ when $F_{\rm sat}\lesssim10-15\%,$ the role of centrals becomes dominant; the $\rvir-$based model predicts that $\mhalo$ increases with $\rhalf$ for centrals, which is reflected in the reversal of the small-scale signal in the bottom left panel. 

The right panels of Figure \ref{fig:lensingprediction} compare the distributions of host halo mass, $\mhost,$ for small and large samples; for centrals, $\mhost=\mhalo,$ while for satellites, $\mhost$ is the virial mass of the parent halo. The peak of each distribution arises from central galaxies, while the long tail at large values of $\mhost$ is due to satellites. At all stellar masses, the distributions in blue peak at slightly larger $\mhost$ relative to the red; this is because the $\rvir-$based model predicts that $\mhalo$ increases with $\rhalf$ for centrals. For $\mstar-$complete samples, this effect only becomes dominant when the satellite fraction drops to sufficiently small values. Qualitatively, these features are completely generic predictions of the $\rvir-$based model; quantitatively, the precise value of $\mstar$ where the effect of centrals begins to dominate $\Delta\Sigma$ depends sensitively on $F_{\rm sat}(\mstar),$ which we have only roughly calibrated in the present work. 

Our methodology is closely aligned with \citet{somerville_etal17}, who studied the empirical modeling features that are necessary to recover the tight scatter in the observed $\mean{\rhalf}{\mstar}$ relation. By building models where $\rhalf$ is set by halo spin $\lambda_{\rm halo}$, it was found in \citet{somerville_etal17} that the level of intrinsic scatter about $\langle\lambda_{\rm halo}\vert\mhalo\rangle$ in dark matter halos is at least as large as the scatter about $\langle\rhalf\vert\mstar\rangle$ seen in observed galaxies. In our approach, the level of scatter is simply a modeling parameter whose fiducial value was motivated by \citet{somerville_etal17}. In ongoing follow-up work discussed in \S\ref{sec:future}, we will systematically test the large-scale structure implications of the assumption that $\rhalf^{\rm disk}\propto\lambda_{\rm halo}$ using forward-modeling methods analogous to those advocated for in \citet{somerville_etal17} and employed here.

\section{Future Directions for Empirical Modeling of Galaxy Size}
\label{sec:future}

This work is intended to serve as a pilot study in which we identify the chief ingredients that influence the $\rhalf-$dependence of galaxy clustering. While our results suggest that sizes of galaxies are linearly correlated with the
virial radius of halos at the moment the halo mass reached its peak, and also that sizes of satellite galaxies were set {\em prior} to infall, our model does not capture galaxy properties and their trends in full quantitative detail.
For example, the clustering ratios of this model shown in Figure \ref{fig:clustering_ratio_upshot} do capture the overall trends, but do not match the observed clustering statistics at the level of reduced $\chi^2\approx 1.$ For present purposes, we have chosen to focus on exploring bracketing cases of qualitative ingredients, rather than fine-tuning the model.

Since the clustering signal is strongly influenced by differences between centrals and satellites, then the satellite fraction $F_{\rm sat}(\mstar)$ plays an important role in $\rhalf-$dependent clustering. Even for {\em fixed} scaling relations $\median{\rhalf^{\rm cens}}{\mhalo}$ and $\median{\rhalf^{\rm sats}}{\mhalo},$ models with different satellite fractions will exhibit distinct $\rhalf-$clustering ratios, because $F_{\rm sat}(\mstar)$ controls the relative weighting of the two populations. An explicit demonstration of this point appears in the Appendix, which repeats the analyses in the main body of the text, but for a subhalo catalog with no orphan correction (and therefore a different satellite fraction).

\begin{figure}
\centering
\includegraphics[width=8cm]{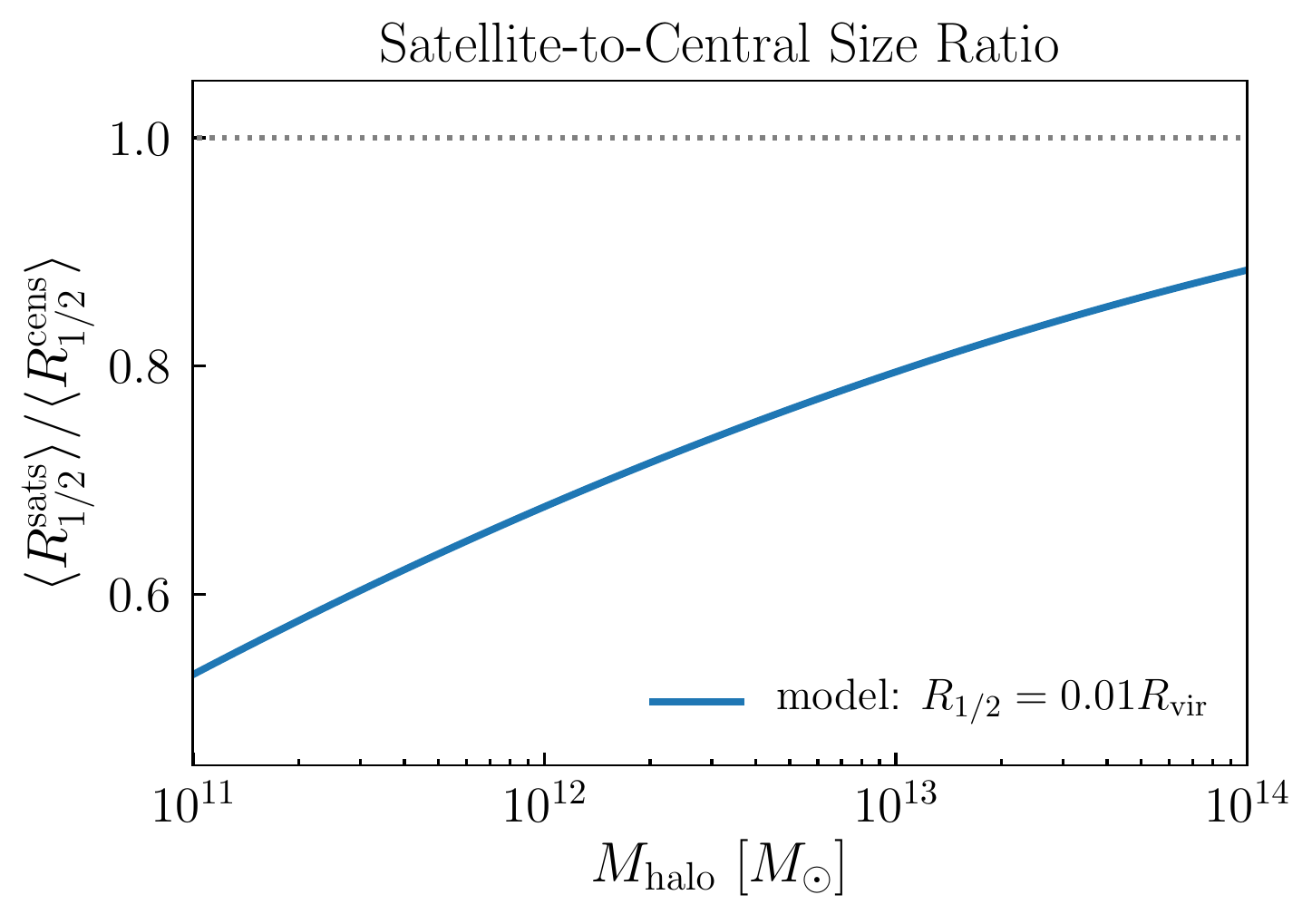}
\caption{
{\bf Scaling relation for hydro sims and SAMs.}
As a function of subhalo mass $\mhalo=\mpeak,$ we show the ratio of mean satellite-to-central size predicted by the $\rvir-$based model. This scaling relation should be a useful guideline for more fine-grained physical models of galaxy size that wish to reproduce the observed $\rhalf-$dependence of galaxy clustering.
}
\label{fig:censatsizeratios}
\end{figure}

On the one hand, this degeneracy with the satellite fraction is unfortunate, because it means galaxy size does not leave a pure and unique signature on $\rhalf-$clustering ratios. However, this can also be viewed as an opportunity to extract tighter constraints on $F_{\rm sat}(\mstar),$ which are sorely needed to discriminate between competing models \citep{watson_conroy13}. Traditional galaxy clustering is already being used to validate and/or fit models of the stellar-to-halo-mass relation \citep[e.g.,][]{leauthaud_etal11,moster_etal10,behroozi13_smhm,lehmann_etal15}. Based on our results, we advocate that empirical models for $\mean{\mstar}{\mhalo}$ be supplemented with additional model ingredients for $\median{\rhalf}{\mhalo},$ and that the parameters of the composite model be {\em jointly} constrained by measurements of the stellar mass function, $\mstar-$dependent clustering, and $\rhalf-$dependent clustering. Because the number of free parameters required to additionally model galaxy size are so few, such analyses will be able to exploit the additional statistical leverage provided by the $\rhalf-$dependence of $\wproj(\rproj)$ to significantly improve constraints on $F_{\rm sat}(\mstar),$ without any appeal to a group-finder.  

In ongoing follow-up work, we are extending the current analysis by jointly modeling stellar mass and size together with a bulge/disk decomposition of model galaxies quantified by ${\rm B/T},$ the fraction of the total stellar mass in the bulge. Forward modeling methods for ${\rm B/T}$ offer potential to significantly improve upon several aspects of current analysis techniques. For example, two-dimensional (projected) sizes of observed galaxies are sometimes de-projected to estimate three-dimensional sizes; such a correction applied to the data requires treating disk-dominated and spheroidal systems separately. The forward modeling approach transforms this correction into parameterized inference: three-dimensional sizes of disk and bulge components are modeled separately and mapped onto dark matter halos; the projected size of model galaxies can then be computed self-consistently with the modeled value of ${\rm B/T},$ permitting a more rigorous, ``apples-to-apples" comparison to observations of projected size. This follow-up analysis will also help robustly quantify the extent to which the ``cause" of satellites being smaller than centrals is due to differences in ${\rm B/T}.$ 

Figure \ref{fig:censatsizeratios} shows that satellites in $\mhalo\approx10^{12}\msun$ halos are $\sim25-40\%$ smaller than their central galaxy counterparts. Since the satellite fraction varies quite significantly with morphology and/or color, we have demonstrated that it is entirely plausible that most of the observed environmental trends of galaxy size can be understood simply in terms of central vs. satellite size differences. These differences are likely not driven by satellite-specific processes inside host halos, but instead are primarily due to the fact that satellites have reached their peak mass at higher redshifts than centrals. While it remains to be seen whether these statistical principles alone are sufficient to capture all the elementary trends that galaxy two-point functions exhibit with galaxy size, morphology and color, one thing is now clear: it is not possible to reliably analyze environmental trends of galaxy $\rhalf$ without properly accounting for satellites.

\section{Conclusions}
\label{sec:conclusion}

We have presented new measurements of the dependence of projected galaxy clustering, $\wproj(\rproj),$ upon galaxy size and stellar mass, and used forward-modeling methods based on {\tt Halotools} to identify the basic ingredients that influence the signal. We conclude with a brief summary of our primary findings:

\ben
\item Small galaxies cluster more strongly than large galaxies of the same stellar mass. Differences between the clustering of small and large galaxies increase on small scales $\rproj\lesssim1\mpc,$ and decrease with increasing stellar mass (see Figure \ref{fig:rvir_only_clustering_absolute}). 
\item The most important factor influencing the clustering signal is the relative size of central and satellite galaxies. The magnitude, scale-dependence, and $\mstar-$dependence of $\rhalf-$dependent clustering provides strong evidence that satellite galaxies are smaller than central galaxies of the same halo mass.
\item A simple empirical model in which $\rhalf$ is proportional to halo $\rvir$ at the time when halo mass reached its maximum reproduces the size-dependence of SDSS galaxy clustering in striking detail (see Figure \ref{fig:clustering_ratio_upshot}). 
\item Models in which $\rhalf$ is regulated by $\mstar,$ rather than $\mhalo,$ are grossly discrepant with the observed $\wproj(\rproj),$ even when accounting for satellite mass stripping (see Figure \ref{fig:mstarmodelclustering}).
\item Taken together, our findings indicate that satellite-specific processes play a sub-dominant role in setting the different sizes of centrals and satellites; the sizes of satellite galaxies instead appear to be largely predetermined at the time of their infall.
\item The $\rvir$-based size model that reproduces the observed $\wproj(\rproj)$ makes an unambiguous  prediction for future observations of galaxy-galaxy lensing: for $\mstar-$selected samples, $\Delta\Sigma$ should be much stronger for small galaxies relative to large (see Figure \ref{fig:lensingprediction}). The strength of the predicted difference in $\Delta\Sigma$ makes this test well within reach of numerous present-day imaging surveys. 
\een

We view the present work as a pilot study that motivates a Bayesian inference program to tightly constrain the galaxy size-halo connection with forward modeling techniques, in direct analogy to the literature on the stellar-to-halo-mass relation.
Our results provide an integral constraint for more complex and fine-grained models of galaxy size, such as semi-analytic models and hydrodynamical simulations. For convenience, Figure \ref{fig:censatsizeratios} provides a simple summary statistic that can be used as a diagnostic for calibrating alternative modeling efforts. 

Our python code has been publicly available on GitHub\footnote{\url{https://github.com/aphearin/galsize_models}} since the inception of the project; we intend for this code to provide a simple means for cosmological surveys to generate synthetic galaxy populations with realistic sizes across the cosmic web. Tabulations of our clustering measurements, as well as catalogs of $z=0$ model galaxies with $\mstar$ and $\rhalf,$ will be made publicly available upon publication; prior to that time, our measurements and mock data are available upon request. 

\section*{Acknowledgments}

APH thanks John Baker for the {\em Toejam \& Earl} soundtrack. Thanks also to Frank van den Bosch, Andrew Zentner, Doug Watson, Simon White, Thorsten Naab, Risa Wechsler and Chihway Chang for thoughtful feedback at various stages of the development of this work, to Faustin Carter and Sebastian Bocquet for sharing their matplotlib expertise, and to Andrew Zentner for Scottish grammar consultation.

We thank the {\tt Astropy} developers for the package-template \citep{astropy}, as well as the developers of {\tt NumPy} \citep{numpy_ndarray}, {\tt SciPy} \citep{scipy}, Jupyter \citep{jupyter}, IPython \citep{ipython}, Matplotlib \citep{matplotlib}, and GitHub for their extremely useful free software. While writing this paper we made extensive use of the Astrophysics Data Service (ADS) and {\tt arXiv} preprint repository.

Work done at Argonne National Laboratory was supported under the DOE contract DE-AC02-06CH11357. This research was supported in part by the National Science Foundation under Grant No. NSF PHY11-25915.  AK was  supported by NSF grant  AST-1412107, and by the Kavli Institute for Cosmological Physics at the University of Chicago through grant PHY-1125897, and an endowment from the Kavli Foundation and its founder Fred Kavli. BPM acknowledges an Emmy Noether grant funded by the Deutsche Forschungsgemeinschaft (DFG, German Research Foundation) -- MO 2979/1-1.

\bibliography{galsize_paper}

\begin{thebibliography}{}

\bibitem[\protect\citeauthoryear{{Abazajian}, {Adelman-McCarthy},
  {Ag{\"u}eros}, {Allam}, {Allende Prieto}, {An}, {Anderson}, {Anderson},
  {Annis}, {Bahcall} \& et al.}{{Abazajian} et~al.}{2009}]{abazajian_etal09}
{Abazajian} K.~N.,  {Adelman-McCarthy} J.~K.,  {Ag{\"u}eros} M.~A.,  {Allam}
  S.~S.,  {Allende Prieto} C.,  {An} D.,  {Anderson} K.~S.~J.,  {Anderson}
  S.~F.,  {Annis} J.,  {Bahcall} N.~A.,    et al. 2009, \apjs, 182, 543

\bibitem[\protect\citeauthoryear{{Ahn}, {Alexandroff}, {Allende Prieto},
  {Anders}, {Anderson}, {Anderton}, {Andrews}, {Aubourg}, {Bailey}, {Bastien}
  \& et al.}{{Ahn} et~al.}{2014}]{ahn_etal14}
{Ahn} C.~P.,  {Alexandroff} R.,  {Allende Prieto} C.,  {Anders} F.,  {Anderson}
  S.~F.,  {Anderton} T.,  {Andrews} B.~H.,  {Aubourg} {\'E}.,  {Bailey} S.,
  {Bastien} F.~A.,    et al. 2014, \apjs, 211, 17

\bibitem[\protect\citeauthoryear{{Astropy Collaboration}, {Robitaille},
  {Tollerud}, {Greenfield}, {Droettboom}, {Bray}, {Aldcroft} et~al.,}{{Astropy
  Collaboration} et~al.}{2013}]{astropy}
{Astropy Collaboration} {Robitaille} T.~P.,  {Tollerud} E.~J.,  {Greenfield}
  P.,  {Droettboom} M.,  {Bray} E.,  {Aldcroft} T.,    et~al., 2013, AAP, 558,
  A33

\bibitem[\protect\citeauthoryear{{Behroozi}, {Conroy} \& {Wechsler}}{{Behroozi}
  et~al.}{2010}]{behroozi_etal10}
{Behroozi} P.~S.,  {Conroy} C.,    {Wechsler} R.~H.,  2010, \apj, 717, 379

\bibitem[\protect\citeauthoryear{{Behroozi}, {Wechsler} \& {Conroy}}{{Behroozi}
  et~al.}{2013}]{behroozi13_smhm}
{Behroozi} P.~S.,  {Wechsler} R.~H.,    {Conroy} C.,  2013, \apj, 770, 57

\bibitem[\protect\citeauthoryear{{Behroozi}, {Wechsler}, {Lu}, {Hahn}, {Busha},
  {Klypin} \& {Primack}}{{Behroozi} et~al.}{2014}]{behroozi_etal14}
{Behroozi} P.~S.,  {Wechsler} R.~H.,  {Lu} Y.,  {Hahn} O.,  {Busha} M.~T.,
  {Klypin} A.,    {Primack} J.~R.,  2014, \apj, 787, 156

\bibitem[\protect\citeauthoryear{{Behroozi}, {Wechsler} \& {Wu}}{{Behroozi}
  et~al.}{2013}]{behroozi12_rockstar}
{Behroozi} P.~S.,  {Wechsler} R.~H.,    {Wu} H.-Y.,  2013, \apj, 762, 109

\bibitem[\protect\citeauthoryear{{Behroozi}, {Wechsler}, {Wu}, {Busha},
  {Klypin} \& {Primack}}{{Behroozi} et~al.}{2013}]{behroozi_etal12b}
{Behroozi} P.~S.,  {Wechsler} R.~H.,  {Wu} H.-Y.,  {Busha} M.~T.,  {Klypin}
  A.~A.,    {Primack} J.~R.,  2013, \apj, 763, 18

\bibitem[\protect\citeauthoryear{{Behroozi}, {Zhu}, {Ferguson}, {Hearin},
  {Lotz}, {Silk}, {Kassin}, {Lu}, {Croton}, {Somerville} \&
  {Watson}}{{Behroozi} et~al.}{2015}]{behroozi_etal15}
{Behroozi} P.~S.,  {Zhu} G.,  {Ferguson} H.~C.,  {Hearin} A.~P.,  {Lotz} J.,
  {Silk} J.,  {Kassin} S.,  {Lu} Y.,  {Croton} D.,  {Somerville} R.~S.,
  {Watson} D.~F.,  2015, \mnras, 450, 1546

\bibitem[\protect\citeauthoryear{{Bernardi}, {Meert}, {Sheth}, {Vikram},
  {Huertas-Company}, {Mei} \& {Shankar}}{{Bernardi}
  et~al.}{2013}]{bernardi_etal13}
{Bernardi} M.,  {Meert} A.,  {Sheth} R.~K.,  {Vikram} V.,  {Huertas-Company}
  M.,  {Mei} S.,    {Shankar} F.,  2013, \mnras, 436, 697

\bibitem[\protect\citeauthoryear{{Bernardi}, {Meert}, {Vikram},
  {Huertas-Company}, {Mei}, {Shankar} \& {Sheth}}{{Bernardi}
  et~al.}{2014}]{bernardi_etal14}
{Bernardi} M.,  {Meert} A.,  {Vikram} V.,  {Huertas-Company} M.,  {Mei} S.,
  {Shankar} F.,    {Sheth} R.~K.,  2014, \mnras, 443, 874

\bibitem[\protect\citeauthoryear{{Bottrell}, {Torrey}, {Simard} \&
  {Ellison}}{{Bottrell} et~al.}{2017}]{bottrell_etal17b}
{Bottrell} C.,  {Torrey} P.,  {Simard} L.,    {Ellison} S.~L.,  2017, \mnras,
  467, 2879

\bibitem[\protect\citeauthoryear{{Brinchmann}, {Charlot}, {White}, {Tremonti},
  {Kauffmann}, {Heckman} \& {Brinkmann}}{{Brinchmann}
  et~al.}{2004}]{brinchmann_etal04}
{Brinchmann} J.,  {Charlot} S.,  {White} S.~D.~M.,  {Tremonti} C.,  {Kauffmann}
  G.,  {Heckman} T.,    {Brinkmann} J.,  2004, \mnras, 351, 1151

\bibitem[\protect\citeauthoryear{{Bryan} \& {Norman}}{{Bryan} \&
  {Norman}}{1998}]{bryan_norman98}
{Bryan} G.~L.,  {Norman} M.~L.,  1998, \apj, 495, 80

\bibitem[\protect\citeauthoryear{{Cacciato}, {van den Bosch}, {More}, {Mo} \&
  {Yang}}{{Cacciato} et~al.}{2013}]{cacciato_etal13}
{Cacciato} M.,  {van den Bosch} F.~C.,  {More} S.,  {Mo} H.,    {Yang} X.,
  2013, \mnras, 430, 767

\bibitem[\protect\citeauthoryear{{Campbell}, {van den Bosch}, {Hearin},
  {Padmanabhan}, {Berlind}, {Mo}, {Tinker} \& {Yang}}{{Campbell}
  et~al.}{2015}]{campbell_etal15}
{Campbell} D.,  {van den Bosch} F.~C.,  {Hearin} A.,  {Padmanabhan} N.,
  {Berlind} A.,  {Mo} H.~J.,  {Tinker} J.,    {Yang} X.,  2015, \mnras, 452,
  444

\bibitem[\protect\citeauthoryear{{Campbell}, {van den Bosch}, {Padmanabhan},
  {Mao}, {Zentner}, {Lange}, {Jiang} \& {Villarreal}}{{Campbell}
  et~al.}{2017}]{campbell_etal17}
{Campbell} D.,  {van den Bosch} F.~C.,  {Padmanabhan} N.,  {Mao} Y.-Y.,
  {Zentner} A.~R.,  {Lange} J.~U.,  {Jiang} F.,    {Villarreal} A.,  2017,
  ArXiv:1705.06347

\bibitem[\protect\citeauthoryear{{Charlton}, {Hudson}, {Balogh} \&
  {Khatri}}{{Charlton} et~al.}{2017}]{charlton_etal17}
{Charlton} P.~J.~L.,  {Hudson} M.~J.,  {Balogh} M.~L.,    {Khatri} S.,  2017,
  \mnras, 472, 2367

\bibitem[\protect\citeauthoryear{{Coil}, {Newman}, {Croton}, {Cooper}, {Davis},
  {Faber}, {Gerke}, {Koo}, {Padmanabhan}, {Wechsler} \& {Weiner}}{{Coil}
  et~al.}{2008}]{coil_etal08}
{Coil} A.~L.,  {Newman} J.~A.,  {Croton} D.,  {Cooper} M.~C.,  {Davis} M.,
  {Faber} S.~M.,  {Gerke} B.~F.,  {Koo} D.~C.,  {Padmanabhan} N.,  {Wechsler}
  R.~H.,    {Weiner} B.~J.,  2008, \apj, 672, 153

\bibitem[\protect\citeauthoryear{{Conroy}, {Wechsler} \& {Kravtsov}}{{Conroy}
  et~al.}{2006}]{conroy_etal06}
{Conroy} C.,  {Wechsler} R.~H.,    {Kravtsov} A.~V.,  2006, \apj, 647, 201

\bibitem[\protect\citeauthoryear{{Croton}}{{Croton}}{2013}]{croton13}
{Croton} D.~J.,  2013, PASA, 30, e052

\bibitem[\protect\citeauthoryear{{Desmond}, {Mao}, {Wechsler}, {Crain} \&
  {Schaye}}{{Desmond} et~al.}{2017}]{desmond_etal17}
{Desmond} H.,  {Mao} Y.-Y.,  {Wechsler} R.~H.,  {Crain} R.~A.,    {Schaye} J.,
  2017, \mnras, 471, L11

\bibitem[\protect\citeauthoryear{{Duarte} \& {Mamon}}{{Duarte} \&
  {Mamon}}{2014}]{duarte_mamon14}
{Duarte} M.,  {Mamon} G.~A.,  2014, \mnras, 440, 1763

\bibitem[\protect\citeauthoryear{{Dutton}, {van den Bosch}, {Dekel} \&
  {Courteau}}{{Dutton} et~al.}{2007}]{dutton_etal08}
{Dutton} A.~A.,  {van den Bosch} F.~C.,  {Dekel} A.,    {Courteau} S.,  2007,
  \apj, 654, 27

\bibitem[\protect\citeauthoryear{{Dutton}, {van den Bosch}, {Faber}, {Simard},
  {Kassin}, {Koo}, {Bundy}, {Huang}, {Weiner}, {Cooper}, {Newman}, {Mozena} \&
  {Koekemoer}}{{Dutton} et~al.}{2011}]{dutton_etal10}
{Dutton} A.~A.,  {van den Bosch} F.~C.,  {Faber} S.~M.,  {Simard} L.,  {Kassin}
  S.~A.,  {Koo} D.~C.,  {Bundy} K.,  {Huang} J.,  {Weiner} B.~J.,  {Cooper}
  M.~C.,  {Newman} J.~A.,  {Mozena} M.,    {Koekemoer} A.~M.,  2011, \mnras,
  410, 1660

\bibitem[\protect\citeauthoryear{{Erben}, {Hildebrandt}, {Miller}
  et~al.,}{{Erben} et~al.}{2013}]{erben_etal13}
{Erben} T.,  {Hildebrandt} H.,  {Miller} L.,    et~al., 2013, \mnras, 433, 2545

\bibitem[\protect\citeauthoryear{{Gao}, {Springel} \& {White}}{{Gao}
  et~al.}{2005}]{gao_white05}
{Gao} L.,  {Springel} V.,    {White} S.~D.~M.,  2005, \mnras, 363, L66

\bibitem[\protect\citeauthoryear{{Guo} \& {White}}{{Guo} \&
  {White}}{2014}]{guo_white13}
{Guo} Q.,  {White} S.,  2014, \mnras, 437, 3228

\bibitem[\protect\citeauthoryear{{Guo}, {White}, {Boylan-Kolchin}, {De Lucia},
  {Kauffmann}, {Lemson}, {Li}, {Springel} \& {Weinmann}}{{Guo}
  et~al.}{2011}]{guo_etal11}
{Guo} Q.,  {White} S.,  {Boylan-Kolchin} M.,  {De Lucia} G.,  {Kauffmann} G.,
  {Lemson} G.,  {Li} C.,  {Springel} V.,    {Weinmann} S.,  2011, \mnras, 413,
  101

\bibitem[\protect\citeauthoryear{{Guo}, {McIntosh}, {Mo}, {Katz}, {Van Den
  Bosch}, {Weinberg}, {Weinmann}, {Pasquali} \& {Yang}}{{Guo}
  et~al.}{2009}]{guo_etal09}
{Guo} Y.,  {McIntosh} D.~H.,  {Mo} H.~J.,  {Katz} N.,  {Van Den Bosch} F.~C.,
  {Weinberg} M.,  {Weinmann} S.~M.,  {Pasquali} A.,    {Yang} X.,  2009,
  \mnras, 398, 1129

\bibitem[\protect\citeauthoryear{{Hearin}, {Campbell}, {Tollerud}, {Behroozi},
  {Diemer}, {Goldbaum}, {Jennings}, {Leauthaud}, {Mao}, {More}, {Parejko},
  {Sinha}, {Sip{\"o}cz} \& {Zentner}}{{Hearin} et~al.}{2017}]{hearin_etal16}
{Hearin} A.~P.,  {Campbell} D.,  {Tollerud} E.,  {Behroozi} P.,  {Diemer} B.,
  {Goldbaum} N.~J.,  {Jennings} E.,  {Leauthaud} A.,  {Mao} Y.-Y.,  {More} S.,
  {Parejko} J.,  {Sinha} M.,  {Sip{\"o}cz} B.,    {Zentner} A.,  2017, \aj,
  154, 190

\bibitem[\protect\citeauthoryear{{Hearin} \& {Watson}}{{Hearin} \&
  {Watson}}{2013}]{hearin_watson13}
{Hearin} A.~P.,  {Watson} D.~F.,  2013, \mnras, 435, 1313

\bibitem[\protect\citeauthoryear{{Heymans}, {Van Waerbeke}, {Miller}
  et~al.,}{{Heymans} et~al.}{2012}]{heymans_etal12}
{Heymans} C.,  {Van Waerbeke} L.,  {Miller} L.,    et~al., 2012, \mnras, 427,
  146

\bibitem[\protect\citeauthoryear{{Hou}, {Lacey} \& {Frenk}}{{Hou}
  et~al.}{2017}]{hou_etal17}
{Hou} J.,  {Lacey} C.~G.,    {Frenk} C.~S.,  2017, ArXiv:1708.02950

\bibitem[\protect\citeauthoryear{{Huang}, {Fall}, {Ferguson}, {van der Wel},
  {Grogin}, {Koekemoer}, {Lee}, {P{\'e}rez-Gonz{\'a}lez} \& {Wuyts}}{{Huang}
  et~al.}{2017}]{huang_etal17}
{Huang} K.-H.,  {Fall} S.~M.,  {Ferguson} H.~C.,  {van der Wel} A.,  {Grogin}
  N.,  {Koekemoer} A.,  {Lee} S.-K.,  {P{\'e}rez-Gonz{\'a}lez} P.~G.,
  {Wuyts} S.,  2017, \apj, 838, 6

\bibitem[\protect\citeauthoryear{{Huang}, {Ho}, {Peng}, {Li} \&
  {Barth}}{{Huang} et~al.}{2013}]{huang_etal13}
{Huang} S.,  {Ho} L.~C.,  {Peng} C.~Y.,  {Li} Z.-Y.,    {Barth} A.~J.,  2013,
  \apj, 766, 47

\bibitem[\protect\citeauthoryear{{Huertas-Company}, {Mei}, {Shankar}, {Delaye},
  {Raichoor}, {Covone}, {Finoguenov}, {Kneib}, {Le} \&
  {Povic}}{{Huertas-Company} et~al.}{2013}]{huertas_company_etal13a}
{Huertas-Company} M.,  {Mei} S.,  {Shankar} F.,  {Delaye} L.,  {Raichoor} A.,
  {Covone} G.,  {Finoguenov} A.,  {Kneib} J.~P.,  {Le} F.~O.,    {Povic} M.,
  2013, \mnras, 428, 1715

\bibitem[\protect\citeauthoryear{{Huertas-Company}, {Shankar}, {Mei},
  {Bernardi}, {Aguerri}, {Meert} \& {Vikram}}{{Huertas-Company}
  et~al.}{2013}]{huertas_company_etal13b}
{Huertas-Company} M.,  {Shankar} F.,  {Mei} S.,  {Bernardi} M.,  {Aguerri}
  J.~A.~L.,  {Meert} A.,    {Vikram} V.,  2013, \apj, 779, 29

\bibitem[\protect\citeauthoryear{Hunter}{Hunter}{2007}]{matplotlib}
Hunter J.~D.,  2007, Computing In Science \& Engineering, 9, 90

\bibitem[\protect\citeauthoryear{{Jiang} \& {van den Bosch}}{{Jiang} \& {van
  den Bosch}}{2014}]{jiang_vdB14}
{Jiang} F.,  {van den Bosch} F.~C.,  2014, ArXiv:1403.6827

\bibitem[\protect\citeauthoryear{Jones, Oliphant, Peterson et~al.,}{Jones
  et~al.}{2016}]{scipy}
Jones E.,  Oliphant T.,  Peterson P.,    et~al., 2001-2016,
  http://www.scipy.org

\bibitem[\protect\citeauthoryear{{Kauffmann}, {Heckman}, {White}
  et~al.,}{{Kauffmann} et~al.}{2003}]{kauffmann_etal03}
{Kauffmann} G.,  {Heckman} T.~M.,  {White} S.~D.~M.,    et~al., 2003, \mnras,
  341, 33

\bibitem[\protect\citeauthoryear{{Kawamata}, {Ishigaki}, {Shimasaku}, {Oguri}
  \& {Ouchi}}{{Kawamata} et~al.}{2015}]{kawamata_etal15}
{Kawamata} R.,  {Ishigaki} M.,  {Shimasaku} K.,  {Oguri} M.,    {Ouchi} M.,
  2015, \apj, 804, 103

\bibitem[\protect\citeauthoryear{{Khochfar} \& {Silk}}{{Khochfar} \&
  {Silk}}{2006}]{khochfar_silk06}
{Khochfar} S.,  {Silk} J.,  2006, \apjl, 648, L21

\bibitem[\protect\citeauthoryear{{Klypin}, {Yepes}, {Gottl{\"o}ber}, {Prada} \&
  {He{\ss}}}{{Klypin} et~al.}{2016}]{klypin_etal16}
{Klypin} A.,  {Yepes} G.,  {Gottl{\"o}ber} S.,  {Prada} F.,    {He{\ss}} S.,
  2016, \mnras, 457, 4340

\bibitem[\protect\citeauthoryear{{Klypin}, {Trujillo-Gomez} \&
  {Primack}}{{Klypin} et~al.}{2011}]{klypin_etal11}
{Klypin} A.~A.,  {Trujillo-Gomez} S.,    {Primack} J.,  2011, \apj, 740, 102

\bibitem[\protect\citeauthoryear{{Kravtsov}}{{Kravtsov}}{2010}]{kravtsov10}
{Kravtsov} A.,  2010, Advances in Astronomy, 2010, 281913

\bibitem[\protect\citeauthoryear{{Kravtsov}, {Vikhlinin} \&
  {Meshscheryakov}}{{Kravtsov} et~al.}{2014}]{kravtsov_etal14}
{Kravtsov} A.,  {Vikhlinin} A.,    {Meshscheryakov} A.,  2014, ArXiv:1401.7329

\bibitem[\protect\citeauthoryear{{Kravtsov}}{{Kravtsov}}{2013}]{kravtsov13}
{Kravtsov} A.~V.,  2013, \apjl, 764, L31

\bibitem[\protect\citeauthoryear{{Kravtsov}, {Berlind}, {Wechsler}, {Klypin},
  {Gottl{\"o}ber}, {Allgood} \& {Primack}}{{Kravtsov}
  et~al.}{2004}]{kravtsov_etal04}
{Kravtsov} A.~V.,  {Berlind} A.~A.,  {Wechsler} R.~H.,  {Klypin} A.~A.,
  {Gottl{\"o}ber} S.,  {Allgood} B.,    {Primack} J.~R.,  2004, \apj, 609, 35

\bibitem[\protect\citeauthoryear{{Landy} \& {Szalay}}{{Landy} \&
  {Szalay}}{1993}]{landy_szalay93}
{Landy} S.~D.,  {Szalay} A.~S.,  1993, \apj, 412, 64

\bibitem[\protect\citeauthoryear{{Lang}, {Wuyts}, {Somerville} et~al.,}{{Lang}
  et~al.}{2014}]{lang_etal14}
{Lang} P.,  {Wuyts} S.,  {Somerville} R.~S.,    et~al., 2014, \apj, 788, 11

\bibitem[\protect\citeauthoryear{{Lange}, {Driver}, {Robotham} et~al.,}{{Lange}
  et~al.}{2015}]{lange_etal15}
{Lange} R.,  {Driver} S.~P.,  {Robotham} A.~S.~G.,    et~al., 2015, \mnras,
  447, 2603

\bibitem[\protect\citeauthoryear{{Leauthaud}, {Tinker}, {Bundy}
  et~al.,}{{Leauthaud} et~al.}{2012}]{leauthaud_etal11}
{Leauthaud} A.,  {Tinker} J.,  {Bundy} K.,    et~al., 2012, \apj, 744, 159

\bibitem[\protect\citeauthoryear{{Lehmann}, {Mao}, {Becker}, {Skillman} \&
  {Wechsler}}{{Lehmann} et~al.}{2017}]{lehmann_etal15}
{Lehmann} B.~V.,  {Mao} Y.-Y.,  {Becker} M.~R.,  {Skillman} S.~W.,
  {Wechsler} R.~H.,  2017, \apj, 834, 37

\bibitem[\protect\citeauthoryear{{Li}, {Kauffmann}, {Jing}, {White},
  {B{\"o}rner} \& {Cheng}}{{Li} et~al.}{2006}]{li_etal06}
{Li} C.,  {Kauffmann} G.,  {Jing} Y.~P.,  {White} S.~D.~M.,  {B{\"o}rner} G.,
   {Cheng} F.~Z.,  2006, \mnras, 368, 21

\bibitem[\protect\citeauthoryear{{Meert}, {Vikram} \& {Bernardi}}{{Meert}
  et~al.}{2013}]{meert_etal13}
{Meert} A.,  {Vikram} V.,    {Bernardi} M.,  2013, \mnras, 433, 1344

\bibitem[\protect\citeauthoryear{{Meert}, {Vikram} \& {Bernardi}}{{Meert}
  et~al.}{2015}]{meert_etal15}
{Meert} A.,  {Vikram} V.,    {Bernardi} M.,  2015, \mnras, 446, 3943

\bibitem[\protect\citeauthoryear{{Moster}, {Naab} \& {White}}{{Moster}
  et~al.}{2013}]{moster_etal13}
{Moster} B.~P.,  {Naab} T.,    {White} S.~D.~M.,  2013, \mnras, 428, 3121

\bibitem[\protect\citeauthoryear{{Moster}, {Somerville}, {Maulbetsch}, {van den
  Bosch}, {Macci{\`o}}, {Naab} \& {Oser}}{{Moster}
  et~al.}{2010}]{moster_etal10}
{Moster} B.~P.,  {Somerville} R.~S.,  {Maulbetsch} C.,  {van den Bosch} F.~C.,
  {Macci{\`o}} A.~V.,  {Naab} T.,    {Oser} L.,  2010, \apj, 710, 903

\bibitem[\protect\citeauthoryear{{Moustakas}, {Coil}, {Aird}, {Blanton},
  {Cool}, {Eisenstein}, {Mendez}, {Wong}, {Zhu} \& {Arnouts}}{{Moustakas}
  et~al.}{2013}]{moustakas_etal13}
{Moustakas} J.,  {Coil} A.~L.,  {Aird} J.,  {Blanton} M.~R.,  {Cool} R.~J.,
  {Eisenstein} D.~J.,  {Mendez} A.~J.,  {Wong} K.~C.,  {Zhu} G.,    {Arnouts}
  S.,  2013, \apj, 767, 50

\bibitem[\protect\citeauthoryear{P\'erez \& Granger}{P\'erez \&
  Granger}{2007}]{ipython}
P\'erez F.,  Granger B.~E.,  2007, Computing in Science and Engineering, 9, 21

\bibitem[\protect\citeauthoryear{{Planck Collaboration}, {Ade}, {Aghanim},
  {Arnaud}, {Ashdown}, {Aumont}, {Baccigalupi}, {Banday}, {Barreiro},
  {Bartlett} \& et al.}{{Planck Collaboration} et~al.}{2016}]{planck15}
{Planck Collaboration} {Ade} P.~A.~R.,  {Aghanim} N.,  {Arnaud} M.,  {Ashdown}
  M.,  {Aumont} J.,  {Baccigalupi} C.,  {Banday} A.~J.,  {Barreiro} R.~B.,
  {Bartlett} J.~G.,    et al. 2016, AAP, 594, A13

\bibitem[\protect\citeauthoryear{{Ragan-Kelley}, {Perez}, {Granger}, {Kluyver},
  {Ivanov}, {Frederic} \& {Bussonier}}{{Ragan-Kelley} et~al.}{2014}]{jupyter}
{Ragan-Kelley} M.,  {Perez} F.,  {Granger} B.,  {Kluyver} T.,  {Ivanov} P.,
  {Frederic} J.,    {Bussonier} M.,  2014, in American Geophysical Union Fall
  Meeting Abstracts {The Jupyter/IPython architecture: a unified view of
  computational research, from interactive exploration to communication and
  publication}

\bibitem[\protect\citeauthoryear{{Reddick}, {Wechsler}, {Tinker} \&
  {Behroozi}}{{Reddick} et~al.}{2013}]{reddick_etal13}
{Reddick} R.~M.,  {Wechsler} R.~H.,  {Tinker} J.~L.,    {Behroozi} P.~S.,
  2013, \apj, 771, 30

\bibitem[\protect\citeauthoryear{{Riebe}, {Partl}, {Enke}, {Forero-Romero},
  {Gottl{\"o}ber}, {Klypin}, {Lemson}, {Prada}, {Primack}, {Steinmetz} \&
  {Turchaninov}}{{Riebe} et~al.}{2013}]{riebe_etal13}
{Riebe} K.,  {Partl} A.~M.,  {Enke} H.,  {Forero-Romero} J.,  {Gottl{\"o}ber}
  S.,  {Klypin} A.,  {Lemson} G.,  {Prada} F.,  {Primack} J.~R.,  {Steinmetz}
  M.,    {Turchaninov} V.,  2013, Astronomische Nachrichten, 334, 691

\bibitem[\protect\citeauthoryear{{Rodr{\'{\i}}guez-Puebla}, {Avila-Reese},
  {Yang}, {Foucaud}, {Drory} \& {Jing}}{{Rodr{\'{\i}}guez-Puebla}
  et~al.}{2015}]{rodriguez_puebla_etal15}
{Rodr{\'{\i}}guez-Puebla} A.,  {Avila-Reese} V.,  {Yang} X.,  {Foucaud} S.,
  {Drory} N.,    {Jing} Y.~P.,  2015, \apj, 799, 130

\bibitem[\protect\citeauthoryear{{Rodr{\'{\i}}guez-Puebla}, {Behroozi},
  {Primack}, {Klypin}, {Lee} \& {Hellinger}}{{Rodr{\'{\i}}guez-Puebla}
  et~al.}{2016}]{rodriguez_puebla16_bolplanck}
{Rodr{\'{\i}}guez-Puebla} A.,  {Behroozi} P.,  {Primack} J.,  {Klypin} A.,
  {Lee} C.,    {Hellinger} D.,  2016, \mnras, 462, 893

\bibitem[\protect\citeauthoryear{{Shankar}, {Mei}, {Huertas-Company}, {Moreno},
  {Fontanot}, {Monaco}, {Bernardi}, {Cattaneo}, {Sheth}, {Licitra}, {Delaye} \&
  {Raichoor}}{{Shankar} et~al.}{2014}]{shankar_etal14}
{Shankar} F.,  {Mei} S.,  {Huertas-Company} M.,  {Moreno} J.,  {Fontanot} F.,
  {Monaco} P.,  {Bernardi} M.,  {Cattaneo} A.,  {Sheth} R.,  {Licitra} R.,
  {Delaye} L.,    {Raichoor} A.,  2014, \mnras, 439, 3189

\bibitem[\protect\citeauthoryear{{Shen}, {Mo}, {White}, {Blanton}, {Kauffmann},
  {Voges}, {Brinkmann} \& {Csabai}}{{Shen} et~al.}{2003}]{shen_etal03}
{Shen} S.,  {Mo} H.~J.,  {White} S.~D.~M.,  {Blanton} M.~R.,  {Kauffmann} G.,
  {Voges} W.,  {Brinkmann} J.,    {Csabai} I.,  2003, \mnras, 343, 978

\bibitem[\protect\citeauthoryear{{Shibuya}, {Ouchi} \& {Harikane}}{{Shibuya}
  et~al.}{2015}]{shibuya_etal15}
{Shibuya} T.,  {Ouchi} M.,    {Harikane} Y.,  2015, \apjs, 219, 15

\bibitem[\protect\citeauthoryear{{Sinha} \& {Garrison}}{{Sinha} \&
  {Garrison}}{2017}]{sinha_etal17}
{Sinha} M.,  {Garrison} L., , 2017, {Corrfunc: Blazing fast correlation
  functions on the CPU}, Astrophysics Source Code Library

\bibitem[\protect\citeauthoryear{{Skibba}, {Coil}, {Mendez}, {Blanton}, {Bray},
  {Cool}, {Eisenstein}, {Guo}, {Miyaji}, {Moustakas} \& {Zhu}}{{Skibba}
  et~al.}{2015}]{skibba_etal15}
{Skibba} R.~A.,  {Coil} A.~L.,  {Mendez} A.~J.,  {Blanton} M.~R.,  {Bray}
  A.~D.,  {Cool} R.~J.,  {Eisenstein} D.~J.,  {Guo} H.,  {Miyaji} T.,
  {Moustakas} J.,    {Zhu} G.,  2015, \apj, 807, 152

\bibitem[\protect\citeauthoryear{{Smith}, {Choi}, {Lee}, {Rhee},
  {Sanchez-Janssen} \& {Yi}}{{Smith} et~al.}{2016}]{smith_etal16}
{Smith} R.,  {Choi} H.,  {Lee} J.,  {Rhee} J.,  {Sanchez-Janssen} R.,    {Yi}
  S.~K.,  2016, \apj, 833, 109

\bibitem[\protect\citeauthoryear{{Somerville}, {Behroozi}, {Pandya}
  et~al.,}{{Somerville} et~al.}{2017}]{somerville_etal17}
{Somerville} R.~S.,  {Behroozi} P.,  {Pandya} V.,    et~al., 2017,
  ArXiv:1701.03526

\bibitem[\protect\citeauthoryear{{Spindler} \& {Wake}}{{Spindler} \&
  {Wake}}{2017}]{spindler_wake17}
{Spindler} A.,  {Wake} D.,  2017, \mnras, 468, 333

\bibitem[\protect\citeauthoryear{{Sunayama}, {Hearin}, {Padmanabhan} \&
  {Leauthaud}}{{Sunayama} et~al.}{2016}]{sunayama_etal16}
{Sunayama} T.,  {Hearin} A.~P.,  {Padmanabhan} N.,    {Leauthaud} A.,  2016,
  \mnras, 458, 1510

\bibitem[\protect\citeauthoryear{{Tasitsiomi}, {Kravtsov}, {Wechsler} \&
  {Primack}}{{Tasitsiomi} et~al.}{2004}]{tasitsiomi_etal04}
{Tasitsiomi} A.,  {Kravtsov} A.~V.,  {Wechsler} R.~H.,    {Primack} J.~R.,
  2004, \apj, 614, 533

\bibitem[\protect\citeauthoryear{{Tinker}, {Leauthaud}, {Bundy}, {George},
  {Behroozi}, {Massey}, {Rhodes} \& {Wechsler}}{{Tinker}
  et~al.}{2013}]{tinker_etal13}
{Tinker} J.~L.,  {Leauthaud} A.,  {Bundy} K.,  {George} M.~R.,  {Behroozi} P.,
  {Massey} R.,  {Rhodes} J.,    {Wechsler} R.~H.,  2013, \apj, 778, 93

\bibitem[\protect\citeauthoryear{{Tinker}, {Weinberg}, {Zheng} \&
  {Zehavi}}{{Tinker} et~al.}{2005}]{tinker_etal05}
{Tinker} J.~L.,  {Weinberg} D.~H.,  {Zheng} Z.,    {Zehavi} I.,  2005, \apj,
  631, 41

\bibitem[\protect\citeauthoryear{{Trujillo}, {Rudnick}, {Rix}, {Labb{\'e}},
  {Franx}, {Daddi}, {van Dokkum}, {F{\"o}rster Schreiber}, {Kuijken},
  {Moorwood}, {R{\"o}ttgering}, {van der Wel}, {van der Werf} \& {van
  Starkenburg}}{{Trujillo} et~al.}{2004}]{trujillo_etal04}
{Trujillo} I.,  {Rudnick} G.,  {Rix} H.-W.,  {Labb{\'e}} I.,  {Franx} M.,
  {Daddi} E.,  {van Dokkum} P.~G.,  {F{\"o}rster Schreiber} N.~M.,  {Kuijken}
  K.,  {Moorwood} A.,  {R{\"o}ttgering} H.,  {van der Wel} A.,  {van der Werf}
  P.,    {van Starkenburg} L.,  2004, \apj, 604, 521

\bibitem[\protect\citeauthoryear{{Vale} \& {Ostriker}}{{Vale} \&
  {Ostriker}}{2004}]{vale_ostriker04}
{Vale} A.,  {Ostriker} J.~P.,  2004, \mnras, 353, 189

\bibitem[\protect\citeauthoryear{{Vale} \& {Ostriker}}{{Vale} \&
  {Ostriker}}{2006}]{vale_ostriker06}
{Vale} A.,  {Ostriker} J.~P.,  2006, \mnras, 371, 1173

\bibitem[\protect\citeauthoryear{{van den Bosch}, {Aquino}, {Yang}, {Mo},
  {Pasquali}, {McIntosh}, {Weinmann} \& {Kang}}{{van den Bosch}
  et~al.}{2008}]{vdB_etal08}
{van den Bosch} F.~C.,  {Aquino} D.,  {Yang} X.,  {Mo} H.~J.,  {Pasquali} A.,
  {McIntosh} D.~H.,  {Weinmann} S.~M.,    {Kang} X.,  2008, \mnras, 387, 79

\bibitem[\protect\citeauthoryear{{Van Der Walt}, {Colbert} \& {Varoquaux}}{{Van
  Der Walt} et~al.}{2011}]{numpy_ndarray}
{Van Der Walt} S.,  {Colbert} S.~C.,    {Varoquaux} G.,  2011, ArXiv:1102.1523

\bibitem[\protect\citeauthoryear{{van der Wel}, {Franx}, {van Dokkum}
  et~al.,}{{van der Wel} et~al.}{2014}]{vanderwel_etal14}
{van der Wel} A.,  {Franx} M.,  {van Dokkum} P.~G.,    et~al., 2014, \apj, 788,
  28

\bibitem[\protect\citeauthoryear{{Vikram}, {Wadadekar}, {Kembhavi} \&
  {Vijayagovindan}}{{Vikram} et~al.}{2010}]{vikram_etal10}
{Vikram} V.,  {Wadadekar} Y.,  {Kembhavi} A.~K.,    {Vijayagovindan} G.~V.,
  2010, \mnras, 409, 1379

\bibitem[\protect\citeauthoryear{{Villarreal}, {Zentner}, {Mao}, {Purcell},
  {van den Bosch}, {Diemer}, {Lange}, {Wang} \& {Campbell}}{{Villarreal}
  et~al.}{2017}]{villarreal_etal17}
{Villarreal} A.~S.,  {Zentner} A.~R.,  {Mao} Y.-Y.,  {Purcell} C.~W.,  {van den
  Bosch} F.~C.,  {Diemer} B.,  {Lange} J.~U.,  {Wang} K.,    {Campbell} D.,
  2017, \mnras, 472, 1088

\bibitem[\protect\citeauthoryear{{Wang}, {Mo} \& {Jing}}{{Wang}
  et~al.}{2009}]{wang_etal09}
{Wang} H.,  {Mo} H.~J.,    {Jing} Y.~P.,  2009, \mnras, 396, 2249

\bibitem[\protect\citeauthoryear{{Wang}, {Li}, {Kauffmann} \& {De
  Lucia}}{{Wang} et~al.}{2007}]{wang_etal07}
{Wang} L.,  {Li} C.,  {Kauffmann} G.,    {De Lucia} G.,  2007, \mnras, 377,
  1419

\bibitem[\protect\citeauthoryear{{Watson}, {Berlind} \& {Zentner}}{{Watson}
  et~al.}{2012}]{watson_etal12}
{Watson} D.~F.,  {Berlind} A.~A.,    {Zentner} A.~R.,  2012, \apj, 754, 90

\bibitem[\protect\citeauthoryear{{Watson} \& {Conroy}}{{Watson} \&
  {Conroy}}{2013}]{watson_conroy13}
{Watson} D.~F.,  {Conroy} C.,  2013, \apj, 772, 139

\bibitem[\protect\citeauthoryear{{Watson}, {Hearin}, {Berlind}, {Becker},
  {Behroozi}, {Skibba}, {Reyes}, {Zentner} \& {van den Bosch}}{{Watson}
  et~al.}{2015}]{watson_etal14}
{Watson} D.~F.,  {Hearin} A.~P.,  {Berlind} A.~A.,  {Becker} M.~R.,  {Behroozi}
  P.~S.,  {Skibba} R.~A.,  {Reyes} R.,  {Zentner} A.~R.,    {van den Bosch}
  F.~C.,  2015, \mnras, 446, 651

\bibitem[\protect\citeauthoryear{{Wechsler}, {Zentner}, {Bullock}, {Kravtsov}
  \& {Allgood}}{{Wechsler} et~al.}{2006}]{wechsler_etal06}
{Wechsler} R.~H.,  {Zentner} A.~R.,  {Bullock} J.~S.,  {Kravtsov} A.~V.,
  {Allgood} B.,  2006, \apj, 652, 71

\bibitem[\protect\citeauthoryear{{Weinmann}, {Kauffmann}, {van den Bosch},
  {Pasquali}, {McIntosh}, {Mo}, {Yang} \& {Guo}}{{Weinmann}
  et~al.}{2009}]{weinmann_etal08}
{Weinmann} S.~M.,  {Kauffmann} G.,  {van den Bosch} F.~C.,  {Pasquali} A.,
  {McIntosh} D.~H.,  {Mo} H.,  {Yang} X.,    {Guo} Y.,  2009, \mnras, 394, 1213

\bibitem[\protect\citeauthoryear{{Yang}, {Mo}, {Jing} \& {van den
  Bosch}}{{Yang} et~al.}{2005}]{yang_etal05b}
{Yang} X.,  {Mo} H.~J.,  {Jing} Y.~P.,    {van den Bosch} F.~C.,  2005, \mnras,
  358, 217

\bibitem[\protect\citeauthoryear{{Yang}, {Mo}, {van den Bosch} \&
  {Jing}}{{Yang} et~al.}{2005}]{yang_etal05a}
{Yang} X.,  {Mo} H.~J.,  {van den Bosch} F.~C.,    {Jing} Y.~P.,  2005, \mnras,
  356, 1293

\bibitem[\protect\citeauthoryear{{Zehavi}, {Zheng}, {Weinberg}
  et~al.,}{{Zehavi} et~al.}{2011}]{zehavi_etal11}
{Zehavi} I.,  {Zheng} Z.,  {Weinberg} D.~H.,    et~al., 2011, \apj, 736, 59

\bibitem[\protect\citeauthoryear{{Zhang} \& {Yang}}{{Zhang} \&
  {Yang}}{2017}]{zhang_yang17}
{Zhang} Y.,  {Yang} X.,  2017, ArXiv:1707.04979

\end{thebibliography}

\newpage
\section*{Appendix: Treatment of Disrupted Subhalos}


Recent results indicate that some portion of subhalos may be artificially disrupted \citep{guo_white13, campbell_etal17}. To address this possibility in our modeling, we use an extension of {\tt Consistent Trees} that models the evolution of subhalos after disruption. The phase space evolution of disrupted subhalos is approximated by following a point mass evolving in the host halo potential according to the orbital parameters of the subhalo at the time of disruption; the evolution of subhalo mass and circular velocity is approximated using the semi-analytic model presented in \citet{jiang_vdB14}. We then use the {\tt extract\_orphans} program in {\tt UniverseMachine} to walk through all the Bolshoi-Planck {\tt hlist} files, yielding the main progenitor information of every subhalo that was ever identified by {\tt Consistent Trees}. 

In our initial application of deconvolution abundance matching we derive the $\mstar-\mpeak$ relation using {\em all} subhalos, including those that were initially considered disrupted. We then apply a selection function to the disrupted subhalos, so that a fraction of these objects will host model galaxies. We refer to this as the {\em orphan selection function}, $\mathcal{F}_{\rm orphan},$ which we consider to be an integral component of our application of abundance matching. Once $\mathcal{F}_{\rm orphan}$ has been applied, the assemblage of calculations described above provides values of $\mpeak,$ $\zpeak,$ and $\mvir$ for disrupted subhalos, which are the only properties needed to apply either the $\rvir-$based or $\mstar-$based size model.

In this appendix, we present two distinct choices for the orphan selection function, described in detail below. For the results in the main body of the text, we have opted for the simplest parameterization we could find that yields reasonably accurate recovery of the galaxy clustering signal observed in SDSS. While a rigorous calibration of $\mathcal{F}_{\rm orphan}$ is beyond the scope of the present work, we point out that a systematic study of subhalo disruption is needed to conduct the program discussed in \S\ref{sec:future}.

In both treatments studied here, we first discard orphan subhalos with $V_{\rm max}/V_{\rm peak}<0.1,$ so that the most drastically disrupted subhalos remain unpopulated with galaxies. For our fiducial model used throughout the main body of the text, $\mathcal{F}_{\rm orphan}$ is defined by randomly selecting half of the orphan subhalos. For the alternative orphan selection function labeled as ``$M_{\rm host}-$dependent orphans" in this appendix, we proceed as follows. We model $\mathcal{F}_{\rm orphan}=\mathcal{F}_{\rm orphan}(M_{\rm peak},M_{\rm host}),$ where $\mpeak$ is the peak mass of the disrupted subhalo, and $\mhost$ is the present-day virial mass of its $z=0$ host halo. For the $\mpeak-$dependence, we select $50\%$ of disrupted subhalos with $\mpeak=10^{11}\msun,$ $0\%$ of subhalos with $\mpeak=10^{13}\msun,$ linearly interpolating in $\log\mpeak$ for intermediate values of $\mpeak.$ At each $\mpeak,$ the selection of disrupted halos is not random; instead, we preferentially select the subhalos with larger $\mhost,$ which we intend to offset the difficulty of subhalo-finding algorithms to identify subhalos with especially small values of  $\mu\equiv\mpeak/\mhost.$

\begin{figure}
\centering
\includegraphics[width=8cm]{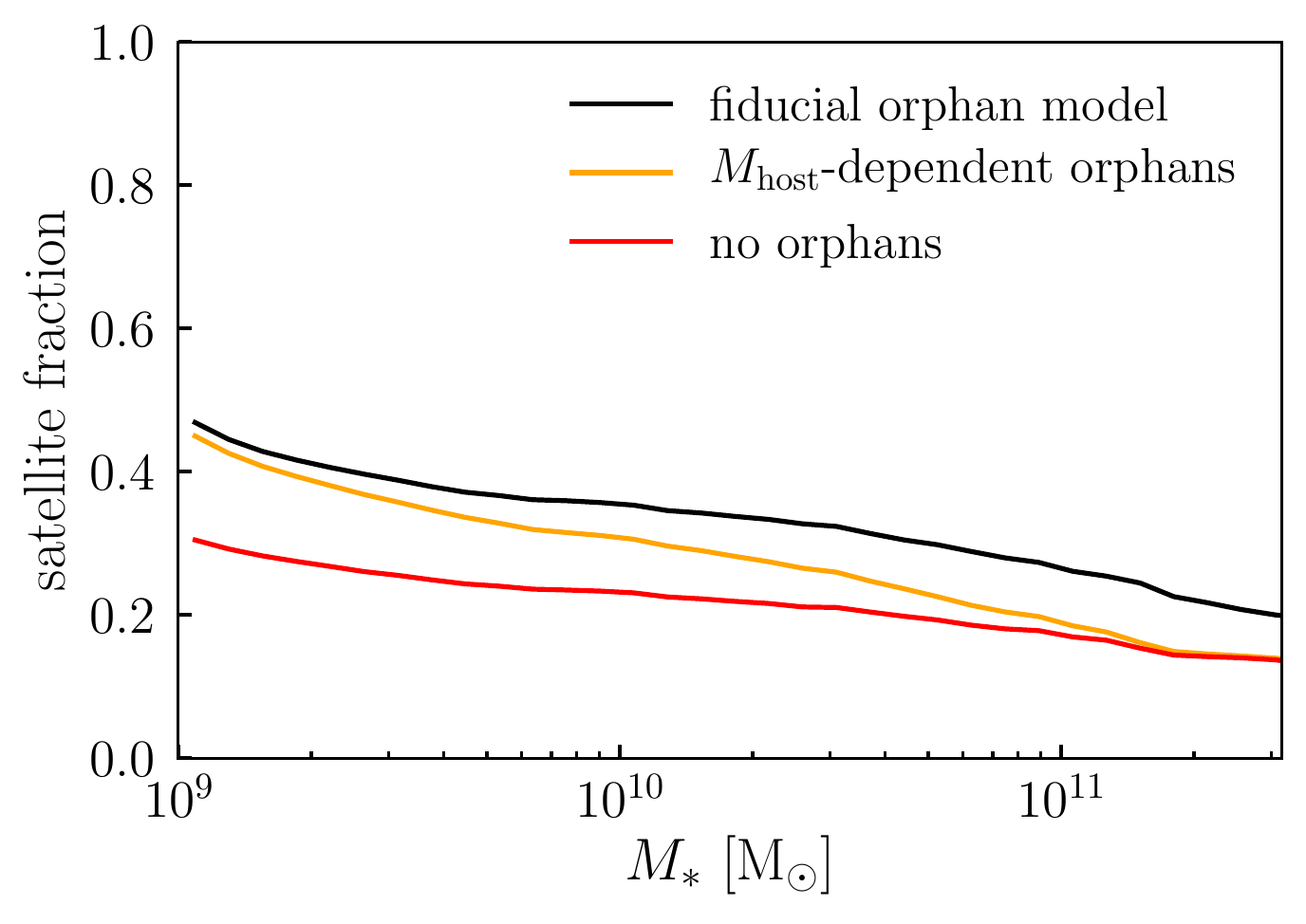}
\caption{
{\bf Impact of orphans on $\fsat.$} The y-axis shows the fraction of galaxies that are not centrals, plotted as a function of stellar mass $\mstar.$ Different curves show results for three different choices for the treatment of disrupted subhalos, as indicated in the legend and described in the text of the appendix. 
}
\label{fig:orphan_satellite_fractions}
\end{figure}

Figure \ref{fig:orphan_satellite_fractions} illustrates the impact of the orphan selection function on the $\mstar-$dependence of the satellite fraction, $\fsat,$ defined as the fraction of galaxies that are not centrals. Both models that include disrupted subhalos receive a significant boost to $\fsat$ due to orphan galaxies. The $\mhost-$dependent model suppresses the selection of disrupted subhalos with larger values of $\mpeak,$ resulting in a milder boost to $\fsat$ at large $\mstar.$ 

\begin{figure*}
\centering
\includegraphics[width=12cm]{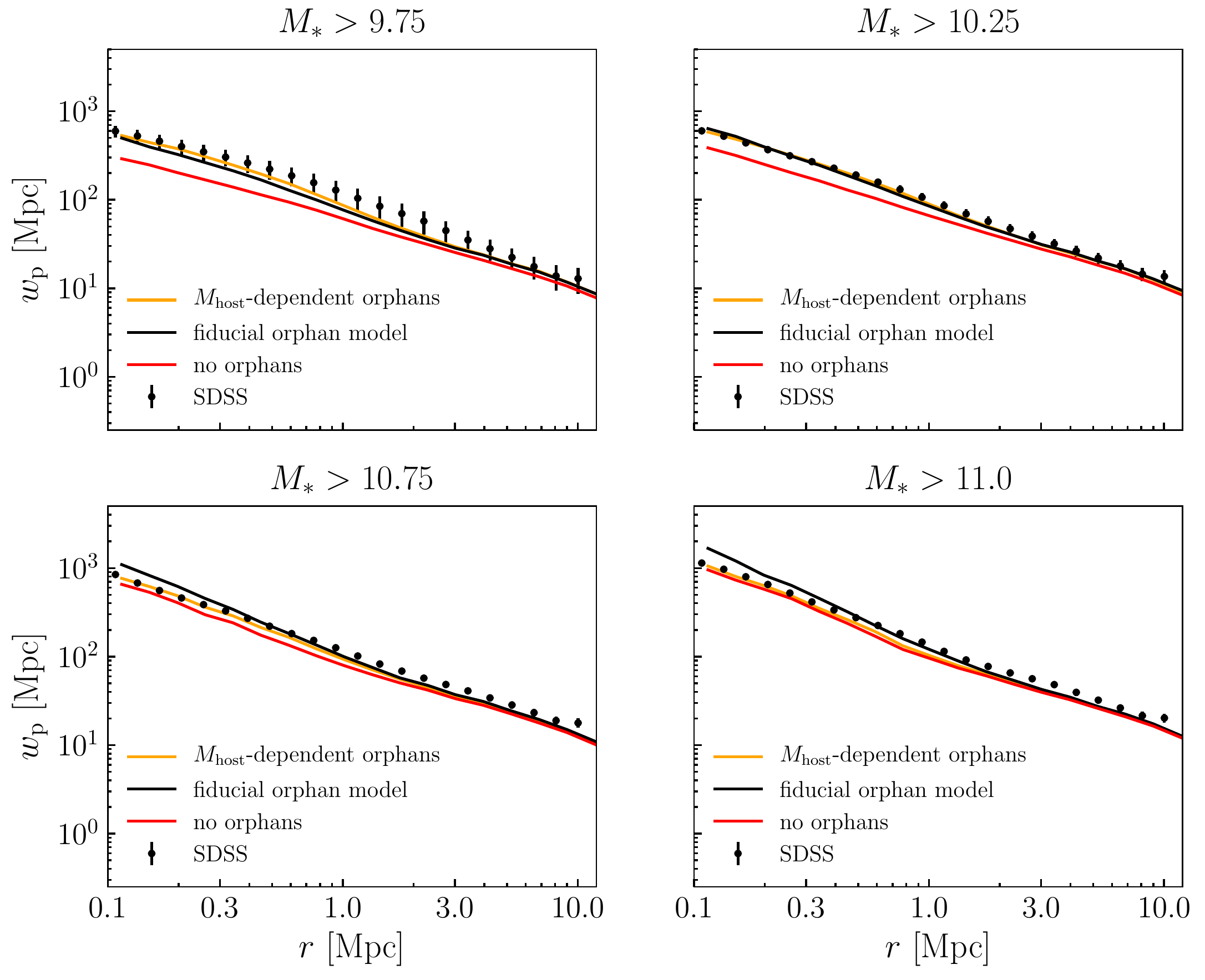}
\caption{
{\bf Impact of orphans on $\mstar-$dependent clustering.}  
Identical to Figure \ref{fig:baseline_sham_clustering}, but including additional curves for the case of $M_{\rm host}-$dependent orphan selection (orange curves), and no orphans at all (red curves). 
}
\label{fig:orphan_baseline_clustering}
\end{figure*}

\begin{figure*}
\centering
\includegraphics[width=12cm]{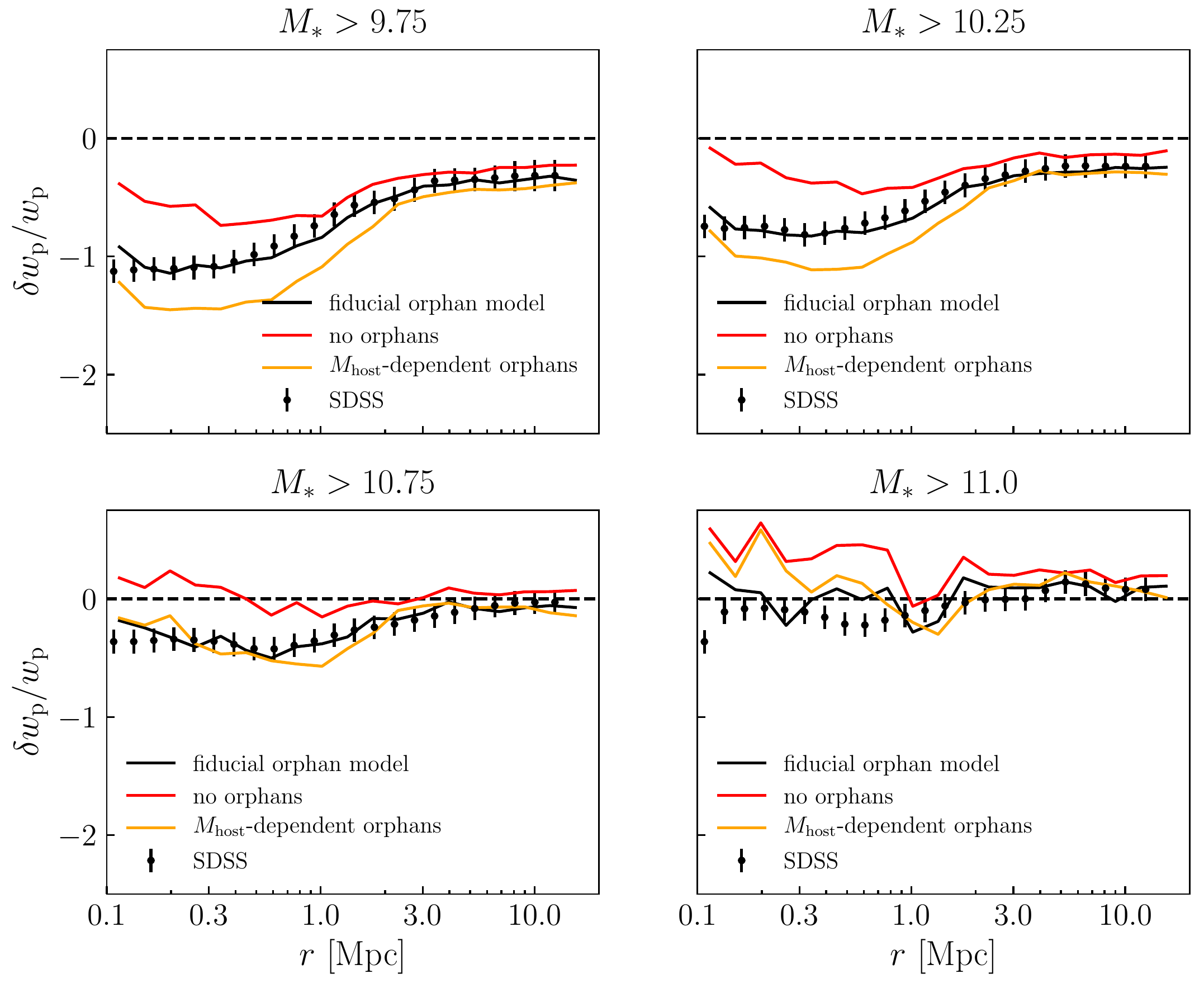}
\caption{
{\bf Impact of orphans on $\rhalf-$dependent clustering.}  
Identical to Figure \ref{fig:clustering_ratio_upshot}, but including additional curves for the case of $M_{\rm host}-$dependent orphan selection (orange curves), and no orphans at all (red curves). 
}
\label{fig:orphan_size_clustering}
\end{figure*}

Figure \ref{fig:orphan_baseline_clustering} shows how the treatment of orphans impacts $\mstar-$dependent clustering. Comparing either the black or orange curves to the red curves shows that including orphans substantially boosts small-scale clustering due to the increased satellite fraction. We show the impact of orphans on $\rhalf-$dependent clustering in Figure \ref{fig:orphan_size_clustering}. As discussed in \S\ref{subsec:censat_sizes}, the value of $\fsat$ modulates the magnitude of the influence of central/satellite size differences on $\wproj;$ including orphans thus boosts the clustering ratios $\delta\wproj/\wproj.$ Because we have not been systematic in the calibration of $\mathcal{F}_{\rm orphan},$ we adopt the simpler, random orphan selection as the fiducial model shown in the main body of the text.


\end{document}